\documentclass[11pt]{article}

\usepackage[dvips]{graphicx}
\usepackage[english]{babel}
\usepackage{psfrag}

\usepackage{cite}

\usepackage{amsmath}

\newcommand{\half}{{\textstyle\frac{1}{2}}}

\newcommand{\gev}{\,\operatorname{GeV}}

\newcommand{\fm}{\,\operatorname{fm}}

\newcommand{\ms}{\mskip 1.5mu}

\newcommand{\msp}{\phantom{-}}

\newcommand{\lsim}{\raisebox{-3pt}{%
    $\,\stackrel{\textstyle<}{\scriptstyle\sim}$\,}}

\newcommand{\mom}[1]{\langle #1 \ms\rangle}

\begin{document}

\begin{flushright}
DESY 07-209 \\
SI-HEP-2007-18 \\
WUB 07-11 \\
\end{flushright}

\begin{center}
\vskip 3.0\baselineskip
{\LARGE \bf Form factors and other measures \\[0.3em]
  of strangeness in the nucleon}
\vskip 4.0\baselineskip
M.~Diehl$^{1}$,
Th.~Feldmann$^{2}$
and P.~Kroll$^{3}$
\vskip \baselineskip
\textit{
  1. Deutsches Elektronen-Synchroton DESY, 22603 Hamburg, Germany \\
  2. Theoretische Physik I, Universit\"at Siegen, Emmy Noether Campus,
  57068 Siegen, Germany \\
  3. Fachbereich Physik, Universit\"at Wuppertal, 42097 Wuppertal,
  Germany}
\vskip 5.0\baselineskip
\textbf{Abstract}\\[0.7\baselineskip]
\parbox{0.9\textwidth}{We discuss the phenomenology of strange-quark
  dynamics in the nucleon, based on experimental and theoretical
  results for electroweak form factors and for parton densities. In
  particular, we construct a model for the generalized parton
  distribution that relates the asymmetry $s(x)-\bar s(x)$ between the
  longitudinal momentum distributions of strange quarks and antiquarks
  with the form factor $F_1^s(t)$, which describes the distribution of
  strangeness in transverse position space.}
\end{center}
\vskip 3.0\baselineskip


\section{Introduction}
\label{sec:intro}

The distribution of strange quarks and antiquarks is a nontrivial
aspect of nucleon structure.  Whereas the presence of these
non-valence degrees of freedom is not surprising, given that gluons
can split into $s\bar{s}$ pairs, their relative abundance compared
with $u\bar{u}$ and $d\bar{d}$ pairs reflects the role of quark masses
in nonperturbative dynamics.  Furthermore, asymmetries in the
distribution of $s$ and $\bar{s}$ are not generated by the simple
splitting $g\to s\bar{s}$ and hence are footprints of more subtle
dynamical mechanisms.  Quantities that have received considerable
attention in the recent literature are form factors of electroweak
currents, which are accessible through parity violation in elastic
lepton-nucleon scattering, and the difference between the momentum
distributions of strange quarks and antiquarks, which has in
particular shown to be relevant for the determination of the weak
mixing angle from deep inelastic neutrino-nucleon scattering
\cite{Davidson:2001ji}.  In the present work we point out
interrelations between different measures of strangeness and connect
two of them quantitatively in a particular model.

A number of quantities related to strangeness in the nucleon are
matrix elements of local operators.  In view of our remarks in the
previous paragraph, it is important to note the behavior of these
operators under charge conjugation.  In particular, the
electromagnetic current is $C$ odd and hence sensitive to the
\emph{difference} between contributions from $s$ and $\bar{s}$.  In
contrast, operators like the axial vector current, the energy-momentum
tensor or the scalar current\footnote{%
  We recall that the scalar current $\bar{s} s$ is relevant in
  connection with the pion-nucleon $\sigma$ term, see
  e.g.~\protect\cite{Sainio:2001bq}.}
are $C$ even and thus \emph{add} the contributions from $s$ and
$\bar{s}$.  Large values of nucleon matrix elements would be more
surprising for $C$ odd operators than for $C$ even ones, since for $C$
odd operators they necessitate important effects beyond simple $g\to
s\bar{s}$ fluctuations.

The unpolarized parton densities $s(x)$, $\bar{s}(x)$ and their
longitudinally polarized counterparts $\Delta s(x)$,
$\Delta\bar{s}(x)$ are expectation values of nonlocal operators and
give the momentum distribution of strange quarks or antiquarks in a
fast moving nucleon.  Specific moments of these distributions are
associated with local operators of definite $C$ parity, as we will
specify shortly.

A suitable framework to discuss relations between various quantities
describing nucleon structure is provided by generalized parton
distributions.  They are matrix elements of the same nonlocal
operators that define the usual parton densities, but taken between
proton states of different momenta.  Throughout this work we consider
these distributions at zero skewness $\xi=0$, and for brevity we will
not display this variable.  For our discussion it is useful to
introduce distributions
\begin{align}
  \label{gpd-def}
H^{\bar{q}}(x,t) &= - H^{q}(-x,t) \,,
&
E^{\bar{q}}(x,t) &= - E^{q}(-x,t) \,,
&
\widetilde{H}^{\bar{q}}(x,t) &= \widetilde{H}^{q}(-x,t) \,,
\end{align}
where the different signs on the r.h.s.\ reflect the different behavior of
vector and axial vector operators under charge conjugation.
$H^q(x,t)$, $E^q(x,t)$ and $\widetilde{H}^q(x,t)$ respectively
correspond to $H^q(x,\xi=0,t)$, $E^q(x,\xi=0,t)$ and
$\widetilde{H}^q(x,\xi=0,t)$ in the notation of
\cite{Diehl:2003ny,Belitsky:2005qn}.
Taking $t=0$ and $x>0$ we obtain the usual quark and antiquark
densities of the proton as
\begin{align}
  \label{forward-lim}
H^q(x,0) &= q(x) \,,  & H^{\bar{q}}(x,0) &= \bar{q}(x) \,,
&
\widetilde{H}^q(x,0) &= \Delta q(x) \,,  &
\widetilde{H}^{\bar{q}}(x,0) &= \Delta\bar{q}(x) \,. \quad
\end{align}
A two-dimensional Fourier transform with respect to $t$ gives the
so-called impact parameter densities
\begin{align}
  \label{impact-def}
q(x,b) &= \int \frac{d^2\boldsymbol{\Delta}}{(2\pi)^2}\;
  e^{-i \boldsymbol{b}\cdot \boldsymbol{\Delta}}\,
  H^q(x,t= -\boldsymbol{\Delta}^2) \,,
&
\bar{q}(x,b) &= \int \frac{d^2\boldsymbol{\Delta}}{(2\pi)^2}\;
  e^{-i \boldsymbol{b}\cdot \boldsymbol{\Delta}}\,
  H^{\bar{q}}(x,t= -\boldsymbol{\Delta}^2) \,,
\end{align}
which specify the spatial distribution of quarks or antiquarks with
longitudinal momentum fraction $x$ in the transverse plane, where the
impact parameter $b$ is the transverse distance of the parton from the
center of the proton \cite{Burkardt:2002hr}.  Impact parameter
densities $\Delta q(x,b)$, $\Delta\bar{q}(x,b)$ for longitudinally
polarized quarks and antiquarks in a longitudinally polarized proton
are obtained from $\widetilde{H}^q(x,t)$,
$\widetilde{H}^{\bar{q}}(x,t)$ in full analogy to \eqref{impact-def}.
The Fourier transforms of $E^{q}(x,t)$, $E^{\bar{q}}(x,t)$ describe
the dependence of the impact parameter distribution of unpolarized
quarks or antiquarks on transverse nucleon polarization
\cite{Burkardt:2002hr}.

The distributions just introduced are connected with the form factors
mentioned above by sum rules, i.e.\ by integrals over the momentum
fraction $x$.  In particular, the lowest moments
\begin{align}
  \label{Dirac}
F_1^s(t) &= \int_{-1}^1 dx\, H^s(x,t) =
  \int_0^1 dx\, \bigl[ H^s(x,t) - H^{\bar{s}}(x,t) \bigr] \,,
\\
  \label{Pauli}
F_2^s(t) &= \int_{-1}^1 dx\, E^s(x,t) =
  \int_0^1 dx\, \bigl[ E^s(x,t) - E^{\bar{s}}(x,t) \bigr]
\end{align}
give the strange Dirac and Pauli form factors, which are defined as
\begin{align}
\langle p(p')|\ms \bar{s} \gamma^\mu s \ms| p(p)\rangle
 = F_1^s(t)\, \bar{u}(p') \gamma^\mu u(p) + F_2^s(t)\,
   \bar{u}(p') \frac{i\sigma^{\mu\nu} (p'-p)_\nu}{2m_p} u(p) \,,
\end{align}
where $t=(p-p')^2$ and $m_p$ is the proton mass.  Their normalization
is
\begin{align}
  \label{F12-norm}
  F_1^s(0) &= 0 \,,  &   F_2^s(0) &= \mu_s \,,
\end{align}
where the first condition reflects that the proton has no net
strangeness, whereas the second condition involves the strangeness
magnetic moment $\mu_s$.  Note that the contributions of $F_1^s$ and
$F_2^s$ to the electromagnetic form factors of proton and neutron
appear with a charge factor $e_s = -1/3$.
The sum rules \eqref{Dirac} and \eqref{Pauli} involve the difference
of quark and antiquark distributions, as it must be for form factors
of the current $\bar{s} \gamma^\mu s$.  Taking the Fourier transform
as in \eqref{impact-def} we see that
\begin{align}
  \label{F1-impact}
\int \frac{d^2\boldsymbol{\Delta}}{(2\pi)^2}\;
  e^{-i \boldsymbol{b}\cdot \boldsymbol{\Delta}}\,
  F_1^s(t= -\boldsymbol{\Delta}^2) 
= \int_0^1 dx\, \bigl[ s(x,b) - \bar{s}(x,b) \bigr]
\end{align}
describes the difference between the transverse spatial distributions
of strange quarks and antiquarks, averaged over their momentum
fraction $x$.  Similarly, the Fourier transform of $F_2^s$ describes
the different dependence of the impact parameter distributions on
transverse nucleon polarization.

Further important moments are
\begin{align}
  \label{A20}
A^s_{2,0}(t) &= \int_{-1}^1 dx\, x H^s(x,t) =
  \int_0^1 dx\, x \bigl[ H^s(x,t) + H^{\bar{s}}(x,t) \bigr] \,,
  \intertext{which is a form factor of the energy-momentum tensor for
    strange quarks, and the strange quark contribution to the axial form
    factor,}
  \label{axial}
F_A^s(t) &= \int_{-1}^1 dx\, \widetilde{H}^s(x,t) =
  \int_0^1 dx\, 
  \bigl[ \widetilde{H}^s(x,t) + \widetilde{H}^{\bar{s}}(x,t) \bigr] \,,
\end{align}
which contributes to elastic lepton-nucleon scattering via $Z$
exchange.  Both form factors belong to charge-conjugation even
currents and thus sum over quark and antiquark distributions.

Using \eqref{forward-lim} we can connect the values of form factors at
$t=0$ with moments of the usual quark and antiquark densities.  In
particular, the first condition in \eqref{F12-norm} is equivalent with
$\mom{s-\bar{s}} = 0$, where we introduced the shorthand notation
\begin{equation}
  \label{mom-def}
\mom{f} = \int_0^1 dx\, f(x) \,.
\end{equation}
In contrast, we have nonzero values for
\begin{align}
A^s_{2,0}(0) &= \mom{x (s+\bar{s})} \,,
&
F_A^s(0) &= \mom{\Delta s + \Delta\bar{s}} \,,
\end{align}
which respectively give the fractional contributions of strange quarks
and antiquarks to the longitudinal momentum and to the spin of the
proton.  There is no analogous relation for the second condition in
\eqref{F12-norm} since $E^{s}(x,t)$ and $E^{\bar{s}}(x,t)$ do not
reduce to any parton density for $t=0$.  Let us however mention that
their Fourier transforms with respect to $t$ satisfy positivity
constraints involving the unpolarized and polarized quark or antiquark
densities \cite{Burkardt:2003ck}.

In order to obtain a quantitative feeling for the role of strange
quarks and antiquarks in the proton, we briefly review in
Sect.~\ref{sec:exp} the current experimental knowledge of the form
factors $F_1^s$ and $F_2^s$ and of strange parton distributions.  In
Sect.~\ref{sec:theo} we mention a number of approaches to calculate
the form factors and the momentum asymmetry $s(x) - \bar{s}(x)$
theoretically, which will indicate dynamical mechanisms that can give
rise to these $C$ odd quantities.
In Sect.~\ref{sec:our-model} we develop a model for $H^s(x,t) -
H^{\bar{s}}(x,t)$ and use it to connect at a quantitative level the
asymmetry $s(x) - \bar{s}(x)$ with the form factor $F_1^s(t)$.
According to \eqref{F1-impact} we thus connect the asymmetry between
strange quark and antiquark distributions in longitudinal momentum
with the one in transverse spatial position.  Our results are
summarized in Sect.~\ref{sec:sum}.


\section{Experimental results for strange form factors and parton
  densities}
\label{sec:exp}

\subsection{Electromagnetic form factors} 
\label{sec:ffs} 

\begin{figure}
\begin{center}
\includegraphics[width=.47\textwidth,bb=97 416 437 730,%
    clip=true]{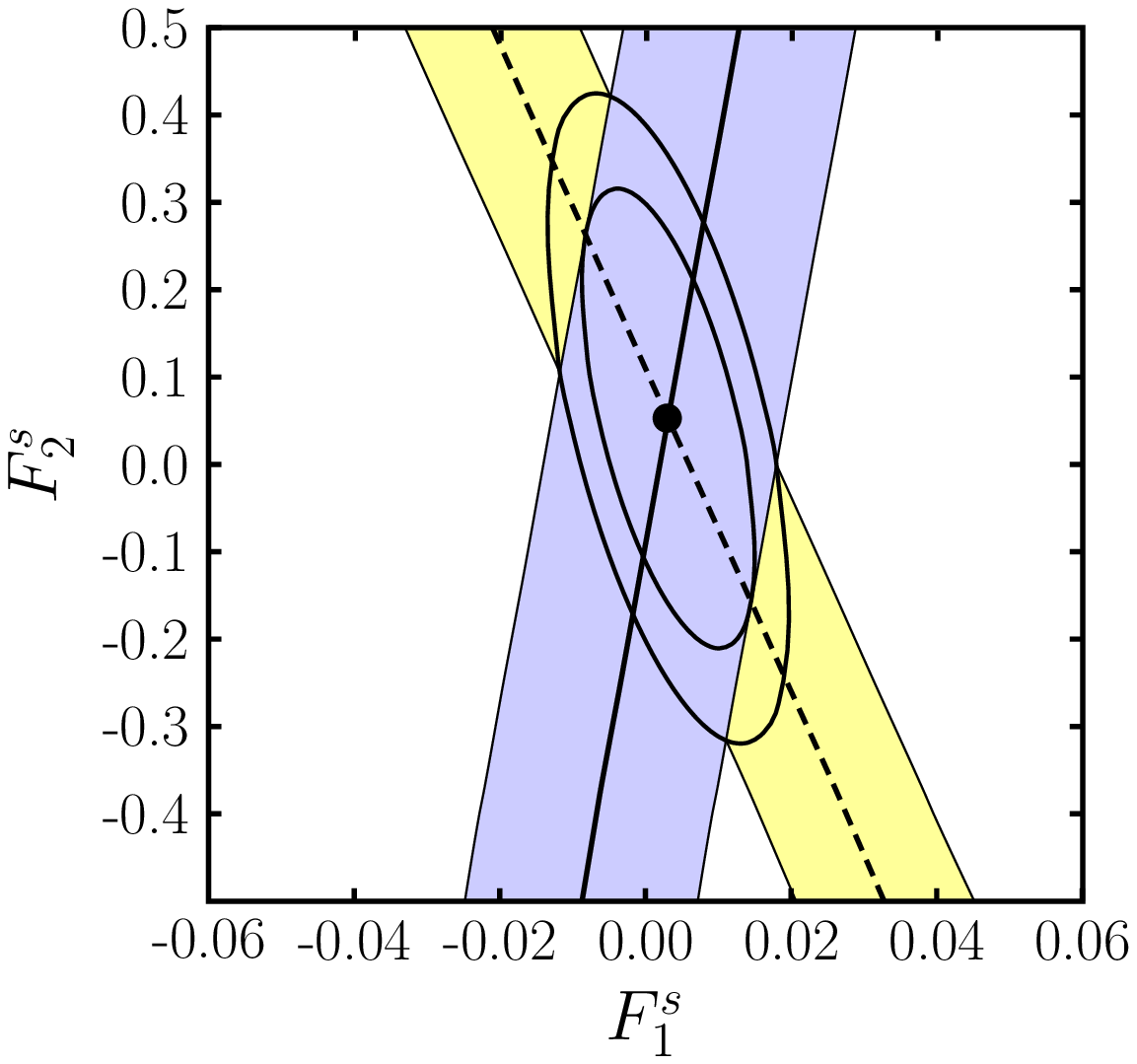}
\end{center}
\caption{\label{fig:ff-data} Results from \protect\cite{Acha:2006my}
  for the strange form factors at $t\simeq -0.1\,\gev^2$.  The dark
  (violet) band represents the result for $G_E^s$, and the light
  (yellow) band the one for $G_E^s+0.09\, G_M^s$.  If one neglects the
  difference of the associated $t$ values one obtains the central
  value given in \protect\eqref{eq:FF-data-1}, which is shown as a
  bullet.  Also shown are the corresponding $1\sigma$ and $2\sigma$
  ellipses.}
\end{figure}

The strange form factors can be extracted from parity violation in
elastic electron scattering on a nucleon
\cite{Spayde:2003nr,Maas:2004ta,%
  Aniol:2004hp,Maas:2004dh,Aniol:2005zf,Aniol:2005zg,%
  Armstrong:2005hs,Acha:2006my}.  Experiments typically measure a
linear combination of the electric and magnetic form factors $G_E^s$
and $G_M^s$ at a low value of the momentum transfer.  This can of
course be converted into a linear combination of $F_1^s$ and $F_2^s$,
using the relations
\begin{align}
  F_1^s &= \frac{G_E^s + \tau G_M^s}{1+\tau}\,, &
  F_2^s &= \frac{G_M^s - G_E^s}{1+\tau} \,, 
\end{align}
where $\tau=-t/4m_p^2$.  Recent experimental results at low $-t$ are
compiled in Table \ref{tab:ff-data}.  Inspection of the table reveals
that only the strange Dirac form factor is fairly well determined,
whereas $F_2^s$ suffers from large uncertainties. It is also evident
that the recent HAPPEX data \cite{Acha:2006my} have significantly
smaller errors than the other measurements.  Unfortunately the two
form factor combinations given in \cite{Acha:2006my} are for different
values of $t$. The determination of the individual form factors
therefore requires an assumption about their $t$ dependence.  A simple
way to proceed is to ignore the difference in $t$ of the two
measurements.  {}From the corresponding two entries in
Table~\ref{tab:ff-data} one then obtains
\begin{align}
  \label{eq:FF-data-1}
F_1^s\bigl( t\simeq -0.1 \gev^2 \bigr) &= 0.003(12) \,, &
F_2^s\bigl( t\simeq -0.1 \gev^2 \bigr) &= 0.05(26)  \,.  
\end{align} 
This result is graphically represented in Fig.~\ref{fig:ff-data}.  The
use of all data near $t=-0.1\,\gev^2$ in Table~\ref{tab:ff-data} does
practically not change $F_1^s$, whereas $F_2^s$ becomes substantially
larger but stays within the error quoted in \eqref{eq:FF-data-1}.

\begin{table}
  \caption{\label{tab:ff-data} Data for the strange form factors at low
    $-t$. Statistical and systematic errors have been added in
    quadrature.  We quote results for $G_M^s$ or $G_E^s + \eta\ms G_M^s$
    and the equivalent ones for $F_1^s + \eta' F_2^s$.}
  \renewcommand{\arraystretch}{1.3} 
\begin{center}
\begin{tabular}{|ccrr|} \hline
 experiment & $-t\, [\gev^2]$ &
 \multicolumn{1}{c}{$G_E^s, G_M^s$} &
 \multicolumn{1}{c|}{$F_1^s, F_2^s$} \\
\hline
SAMPLE \protect\cite{Spayde:2003nr} & 0.100
 & $G_M^s= \msp0.37(34)\phantom{0}$
 & $F_1^s+F_2^s= \msp 0.37(34)\phantom{0}$ \\[0.2em]
A4 \protect\cite{Maas:2004ta} & 0.23\phantom{0}
 &   $G_E^s+0.225\, G_M^s= \msp 0.039(34)$
 & ~~$F_1^s+0.130\, F_2^s= \msp 0.032(28)$ \\[0.2em]
HAPPEX \protect\cite{Aniol:2004hp} & 0.477
 & $G_E^s+0.392\, G_M^s= \msp 0.014(22)$
 & $F_1^s+0.184\, F_2^s= \msp 0.010(16)$ \\[0.2em]
A4 \protect\cite{Maas:2004dh} & 0.108
 & $G_E^s+0.106\, G_M^s= \msp 0.071(36)$
 & $F_1^s+0.068\, F_2^s= \msp 0.064(33)$ \\[0.2em]
HAPPEX \protect\cite{Aniol:2005zf} & 0.091
 & $G_E^s= -0.038(43)$
 & $F_1^s-0.026\, F_2^s= -0.038(43)$ \\[0.2em]
HAPPEX \protect\cite{Aniol:2005zg} & 0.099
 & $G_E^s+0.080\, G_M^s= \msp 0.030(28)$
 & $F_1^s+0.048\, F_2^s= \msp 0.028(26)$ \\[0.2em]
HAPPEX \protect\cite{Acha:2006my} & 0.077
 & $G_E^s= \msp 0.002(16)$
 & $F_1^s-0.022\, F_2^s= \msp 0.002(16)$ \\[0.2em]
HAPPEX \protect\cite{Acha:2006my} & 0.109
 & $G_E^s+0.090\, G_M^s= \msp 0.007(13)$
 & $F_1^s+0.054\, F_2^s= \msp 0.006(12)$ \\[0.2em]
\hline
\end{tabular} 
\end{center} 
\end{table} 

An alternative method has been used in \cite{Young:2006jc}, where a
parameterization of the small $t$ behavior of the strange form factors
was fitted to all data below $-t= 0.3 \gev^2$.  With the results
updated in \cite{Young:2007zs}, the authors of this study obtain
\cite{Thomas:2007pc}
\begin{align}
  \label{eq:FF-data-2}
  F_1^s(t)&= -t \times 0.02(11) \gev^{-2} \,, &
  F_2^s(t)&= -0.01(25) \,, 
\end{align}
for $-t \le 0.3 \gev^2$, which at $-t = 0.1 \gev^2$ is fully compatible
with our simple estimate \eqref{eq:FF-data-1}.  The analysis
in \cite{Young:2006jc,Young:2007zs} includes the data of the G0
Collaboration \cite{Armstrong:2005hs} with their very fine binning in
$t$, which we have not listed in Table \ref{tab:ff-data}.

We remark that the experimental values quoted here are subject to
   theoretical uncertainties due to the effects of two-photon
   and of $\gamma Z$ exchange, which have been discussed in 
   \cite{Afanasev:2005ex,Zhou:2007hr,Tjon:2007wx} and may not be negligible.

\subsection{Unpolarized parton densities}
\label{sec:pdfs}

The determination of parton densities (PDFs) from unpolarized hard
scattering processes has made significant progress in the recent
decade, in particular thanks to data with high precision and a large
kinematical reach from HERA and from the Tevatron.  The knowledge of
strange distributions is much less advanced, because the observables
that dominate global fits of parton densities have little sensitivity
to $s$ or $\bar{s}$.  This holds in particular for inclusive deep
inelastic scattering (DIS) in kinematics where photon exchange is
dominant.  More specific constraints on $s$ and $\bar{s}$
distributions come from fixed-target DIS experiments with $\nu$ and
$\bar{\nu}$ beams.  Thanks to such measurements, there have recently
been dedicated attempts to determine $s(x)$ and $\bar{s}(x)$ without
strong assumptions on their relation with the light sea quark
distributions $\bar{u}(x)$ and $\bar{d}(x)$.

\begin{figure}
\begin{center}
  \psfrag{xx}[c][c]{\small $x$}
  \psfrag{tt}[c][c]{\small $x (s+\bar{s})$}
  \psfrag{bb}[r][c]{\footnotesize CTEQ6.5S$0$}
  \psfrag{dd}[r][c]{\footnotesize Alekhin 06}
  \psfrag{hh}[r][c]{\footnotesize MRST 2006}
\includegraphics[width=0.49\textwidth,%
  bb=80 50 385 295]{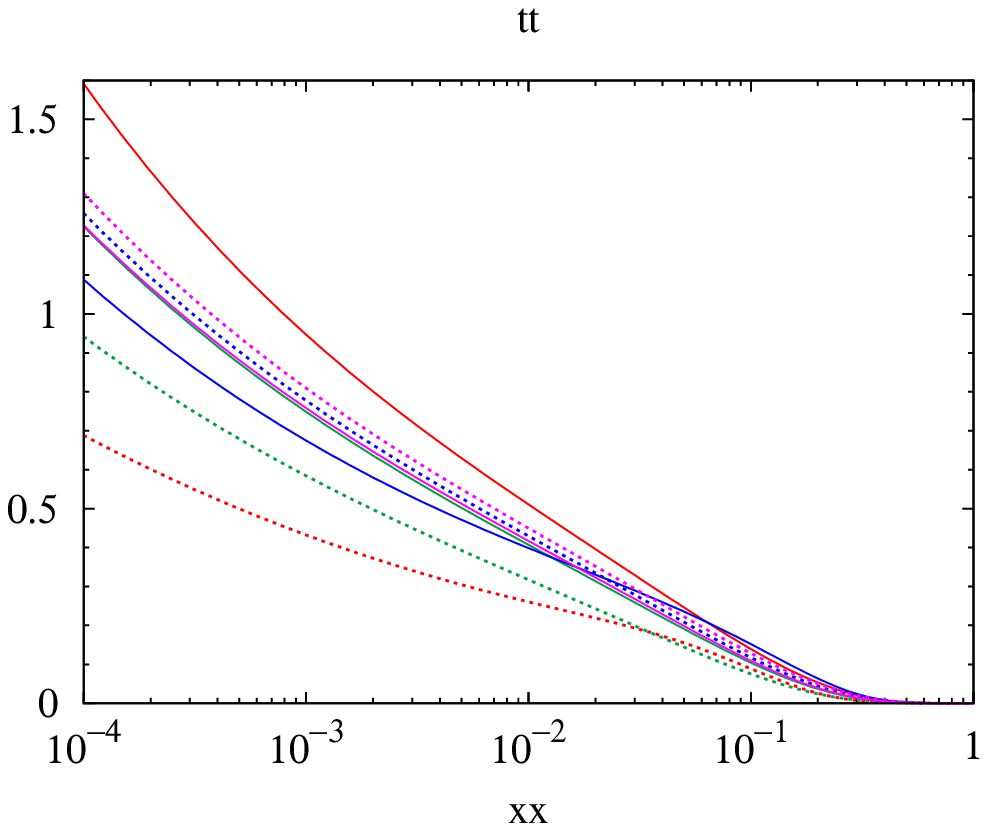}
\includegraphics[width=0.49\textwidth,%
  bb=80 50 385 295]{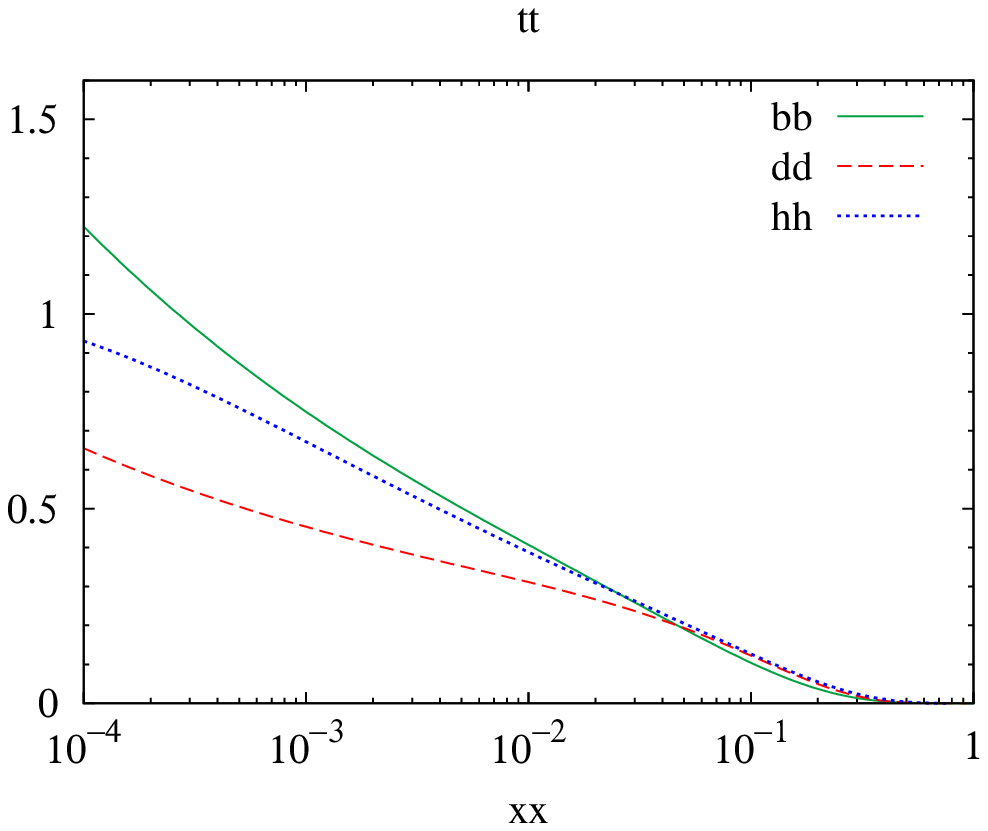}
\\[0.7em]
  \psfrag{tt}[c][c]{\small $x (\bar{u}+\bar{d})$}
\includegraphics[width=0.49\textwidth,%
  bb=80 50 385 295]{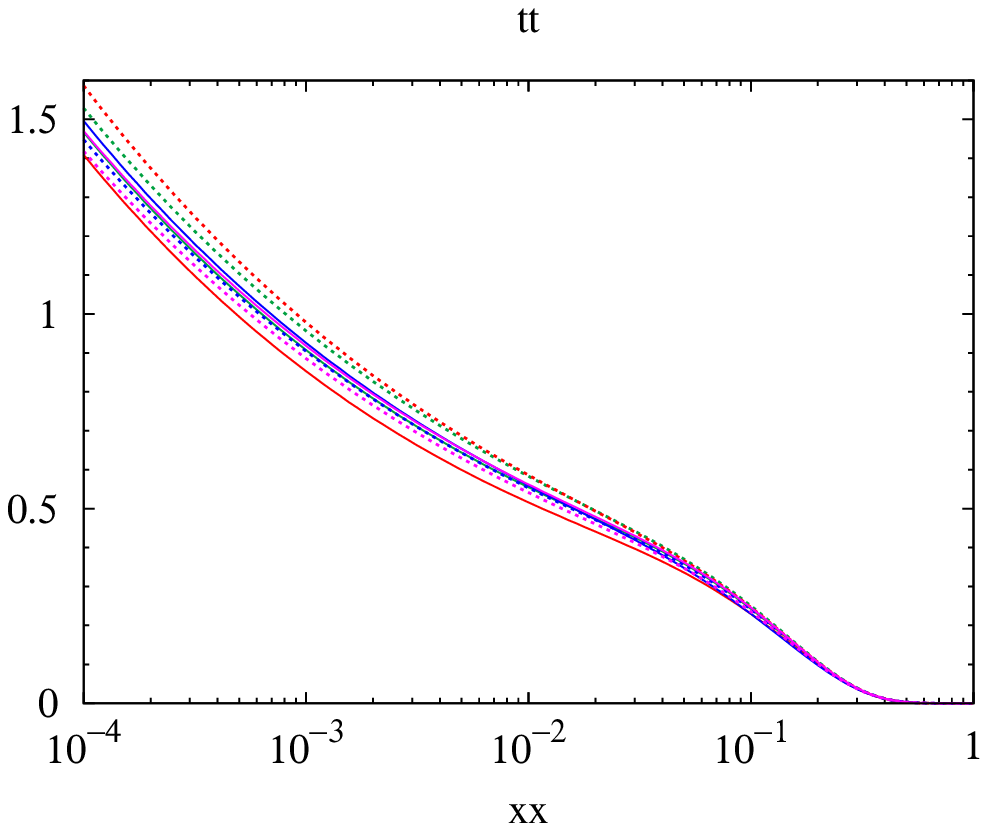}
\includegraphics[width=0.49\textwidth,%
  bb=80 50 385 295]{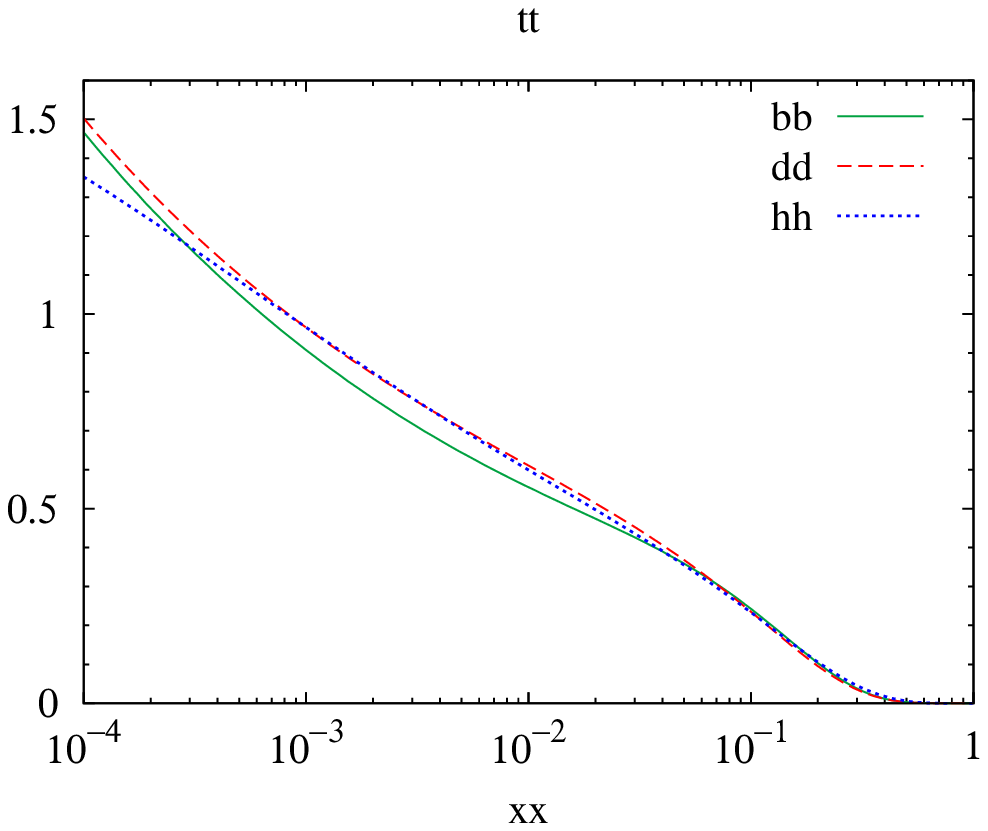}
\\[0.7em]
  \psfrag{tt}[c][c]{\small $x (\bar{u}-\bar{d})$}
\includegraphics[width=0.49\textwidth,%
  bb=80 50 385 295]{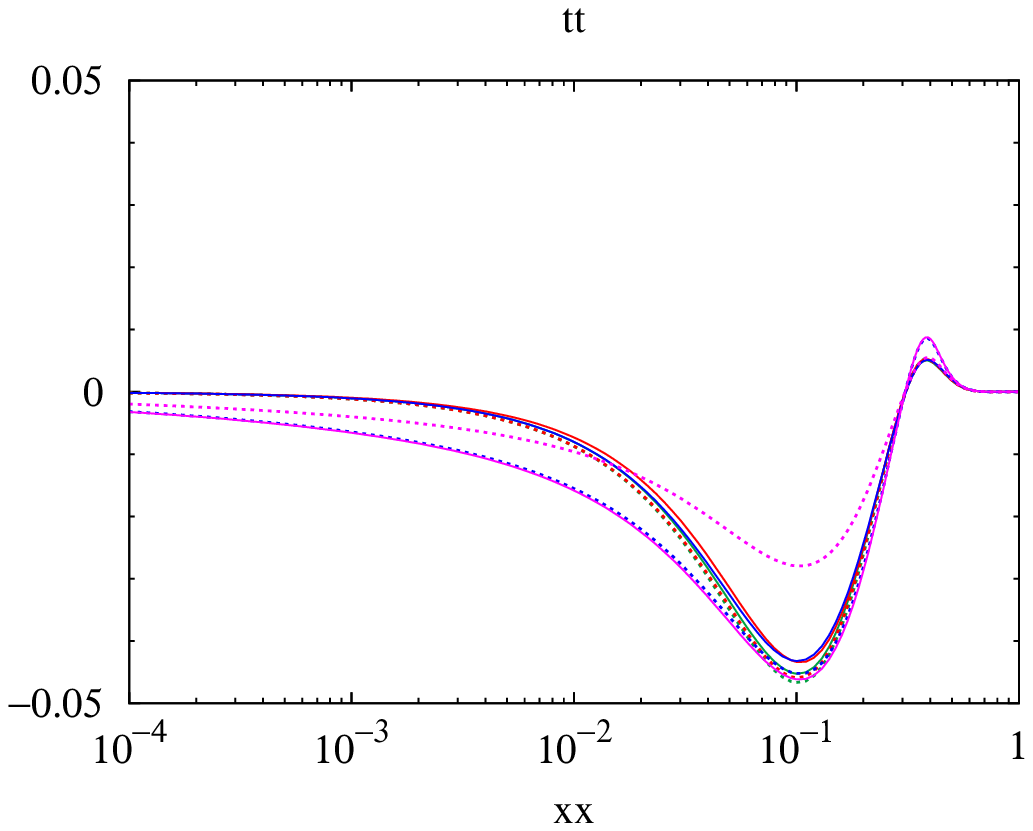}
\includegraphics[width=0.49\textwidth,%
  bb=80 50 385 295]{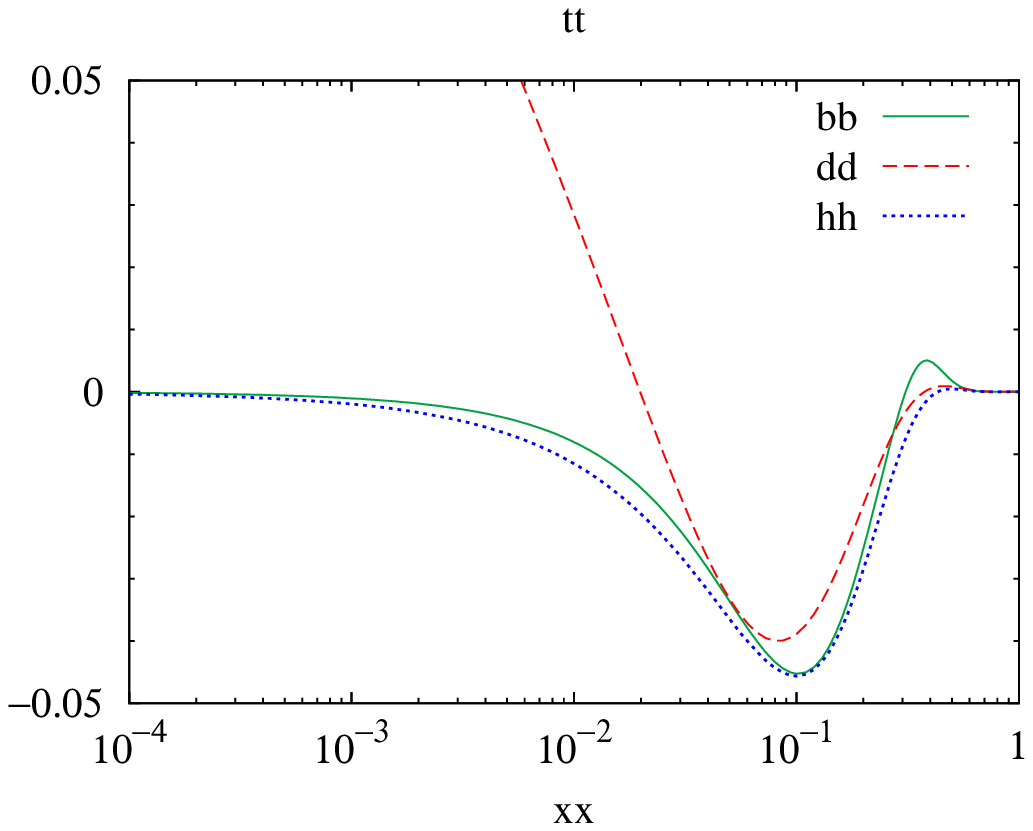}
\end{center}
\caption{\label{fig:pdfs} Different combinations of parton densities
  at $\mu= 2\gev$.  Left: comparison of the different sets from
  CTEQ6.5S \protect\cite{Lai:2007dq}.  Right: comparison of set 0 from
  CTEQ6.5S \protect\cite{Lai:2007dq} with Alekhin 06
  \protect\cite{Alekhin:2006zm} and MRST 2006
  \protect\cite{Martin:2007bv}.}
\end{figure}

In Table~\ref{tab:moments} we give the moments $\mom{x (s+\bar{s})}$
obtained in recent PDF extractions.  The study in \cite{Lai:2007dq}
was dedicated to strangeness and explored a number of fits at NLO in
$\alpha_s$.  The extractions in \cite{Alekhin:2006zm,Martin:2007bv}
were performed at NNLO.  The table also gives the moment $\mom{x
  (\bar{u}+\bar{d})}$, whose values range from $6.21\times 10^{-2}$ to
$6.79\times 10^{-2}$ and thus show a much smaller spread than for
$\mom{x (s+\bar{s})}$.  The ratio of $\mom{x (s+\bar{s})}$ and $\mom{x
  (\bar{u}+\bar{d})}$ is between $0.36$ and $0.72$ and quantifies the
suppression of strangeness in the light quark sea.  We furthermore
show the flavor asymmetry $\mom{x (\bar{u}-\bar{d})}$, which like
$\mom{x (s-\bar{s})}$ is not generated by simple $g\to q\bar{q}$
splitting and hence requires more subtle dynamics in order to be
nonzero.  The ratio of $\mom{x (\bar{u}-\bar{d})}$ and $\mom{x
  (\bar{u}+\bar{d})}$ varies between $-7\%$ and $-14\%$.  The parton
densities corresponding to the entries in the table are plotted in
Fig.~\ref{fig:pdfs}. We note that there are no experimental
constraints on $\bar{u}(x)-\bar{d}(x)$ at small $x$, so that the large
difference between the results of \cite{Alekhin:2006zm} and
\cite{Lai:2007dq,Martin:2007bv} for this combination of densities is a
consequence of the different functional forms assumed in the fits.

\begin{table}
\renewcommand{\arraystretch}{1.2}
\begin{center}
\begin{tabular}{|c|cccccccc|} \hline
     & \multicolumn{8}{c|}{CTEQ6.5S \protect\cite{Lai:2007dq}} \\
 set & $\msp 0$ & $\msp 1$ & $\msp 2$ & $\msp 3$
     & $\msp 4$ & $-0$ & $-1$ & $-2$ \\ \hline
$\mom{x (s+\bar{s})}$
  & $\msp 3.35$ & $\msp 2.46$ & $\msp 4.44$ & $\msp 2.45$
  & $\msp 4.30$ & $\msp 3.72$ & $\msp 3.48$ & $\msp 4.04$ \\
$\mom{x (\bar{u}+\bar{d})}$
  & $\msp 6.55$ & $\msp 6.79$ & $\msp 6.21$ & $\msp 6.74$
  & $\msp 6.35$ & $\msp 6.54$ & $\msp 6.65$ & $\msp 6.37$ \\
$\mom{x (\bar{u}-\bar{d})}$
  & $-0.72$ & $-0.75$ & $-0.69$ & $-0.74$
  & $-0.69$ & $-0.74$ & $-0.76$ & $-0.45$ \\ \hline
\end{tabular} 
\end{center}
\vspace{0.1em}
\begin{center}
\parbox{0.46\textwidth}{
\caption{\label{tab:moments} Results of different PDF fits for the
  moments $\mom{x (s+\bar{s})}$, $\mom{x (\bar{u}+\bar{d})}$ and
  $\mom{x (\bar{u}-\bar{d})}$.  All numbers are given in units of
  $10^{-2}$ and refer to the scale $\mu=2 \gev$ in the
  $\overline{\mathrm{MS}}$ scheme.}
}
\hspace{0.9em}
\begin{tabular}{|c|cc|} \hline
 & Alekhin 06 \protect\cite{Alekhin:2006zm}
 & MRST 2006 \protect\cite{Martin:2007bv} \\ \hline
$\mom{x (s+\bar{s})}$       & $\msp 3.40$ & $\msp 3.89$ \\
$\mom{x (\bar{u}+\bar{d})}$ & $\msp 6.56$ & $\msp 6.77$ \\
$\mom{x (\bar{u}-\bar{d})}$ & $-0.56$     & $-0.92$ \\ \hline
\end{tabular} 
\end{center}
\end{table}


\begin{figure}
\begin{center}
  \psfrag{xx}[c][c]{\small $x$}
  \psfrag{tt}[c][c]{\small $x (s-\bar{s})$}
  \psfrag{bb}[l][c]{\footnotesize set $-2$}
  \psfrag{bd}[l][c]{\footnotesize \phantom{set} $-0$}
  \psfrag{dd}[l][c]{\footnotesize \phantom{set} $-1$}
\includegraphics[width=0.49\textwidth,%
  bb=80 50 385 295]{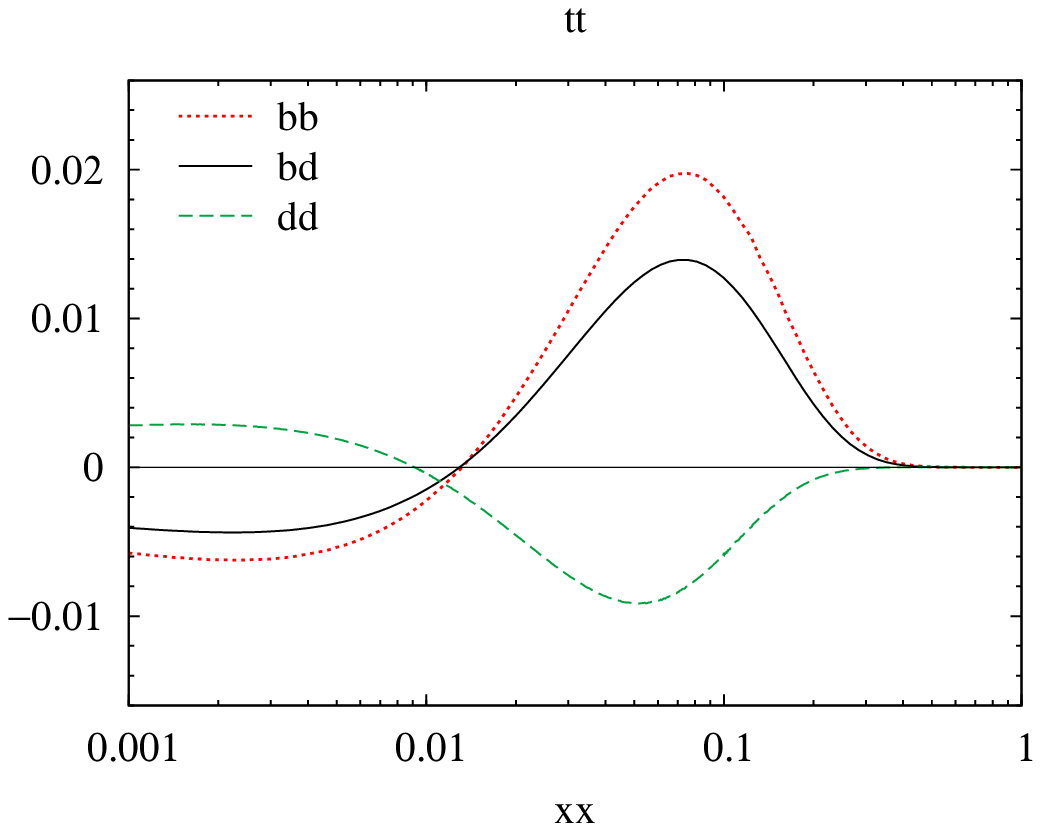}
  \psfrag{bb}[l][c]{\footnotesize B$+$}
  \psfrag{bd}[l][c]{\footnotesize B}
  \psfrag{dd}[l][c]{\footnotesize B$-$}
  \psfrag{hh}[l][c]{\footnotesize C}
\includegraphics[width=0.49\textwidth,%
  bb=80 50 385 295]{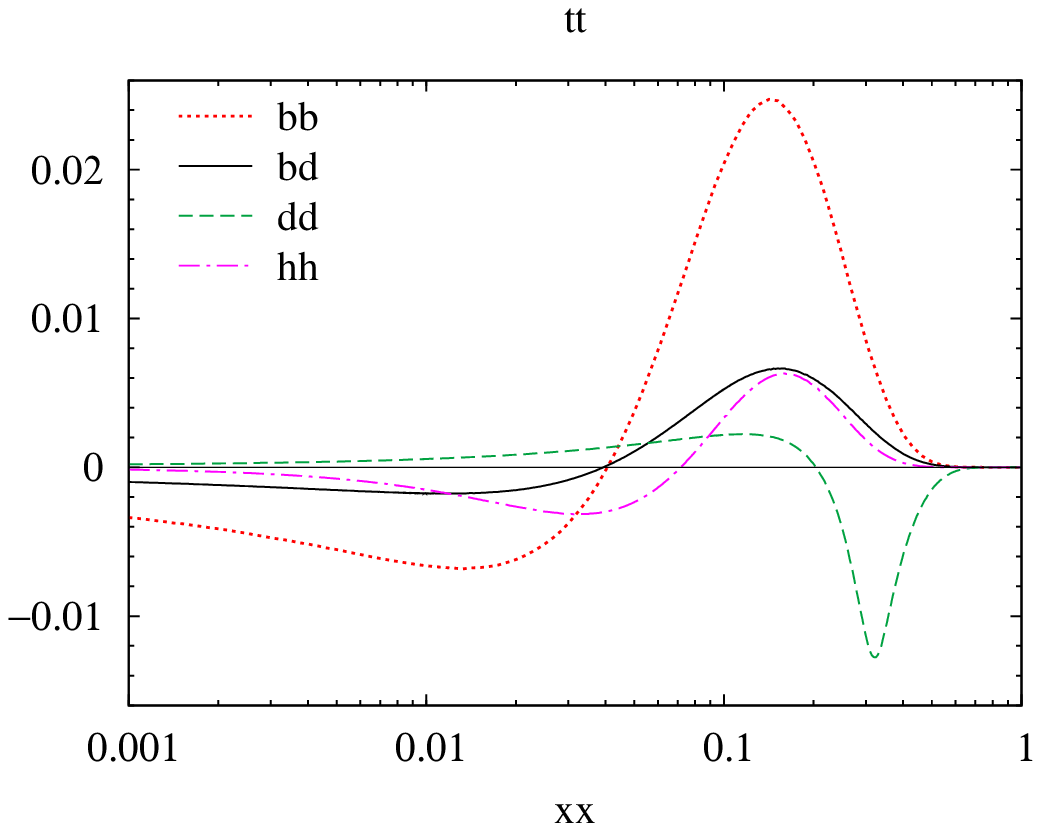}
\end{center}
\caption{\label{fig:s-sbar} The strangeness asymmetry distribution at
  scale $\mu= 2\gev$.  The left panel shows the fits of the CTEQ6.5S
  analysis \protect\cite{Lai:2007dq} and the right panel those of
  CTEQ6 \protect\cite{Olness:2003wz}.  Sets B and C correspond to
  different assumptions on the small-$x$ behavior of
  $s(x)-\bar{s}(x)$, whereas sets $-1$ and B$-$ ($-2$ and B$+$) have
  been chosen to minimize (maximize) the moment $\mom{x
    (s-\bar{s})}$.}
\end{figure}

The CTEQ6.5S study \cite{Lai:2007dq} also performed fits where $s(x)$
and $\bar{s}(x)$ were allowed to be different.  An essential input for
constraining this difference are the CCFR and NuTeV data for dimuon
production in $\nu$ and $\bar{\nu}$ DIS \cite{Goncharov:2001qe}.  The
parameterization in \cite{Lai:2007dq} was chosen such that
$s(x)-\bar{s}(x)$ has precisely one zero crossing.  The resulting
momentum asymmetry at scale $\mu= 2\gev$ was found to be
\begin{equation} 
\mom{x (s-\bar{s})} =
  \begin{cases}
    \msp 2.0\phantom{0} \times 10^{-3} & ~~ (\text{set $-0$}) \,, \\
    -0.94 \times 10^{-3} & ~~ (\text{set $-1$}) \,, \\
    \msp 2.9\phantom{0} \times 10^{-3} & ~~ (\text{set $-2$}) \,,
  \end{cases} 
\end{equation} 
where set $-0$ corresponds to the best fit, whereas sets $-1$ and $-2$
were chosen to give the smallest and largest values of $\mom{x
  (s-\bar{s})}$, respectively.  The ratio of $\mom{x (s-\bar{s})}$ and
$\mom{x (s+\bar{s})}$ in the three fits has respective values $5.4\%$,
$-2.7\%$ and $7.2\%$, which is somewhat smaller in size than the ratio
of $\mom{x (\bar{u}-\bar{d})}$ and $\mom{x (\bar{u}+\bar{d})}$ in the
non-strange sector.  The left panel in Fig.~\ref{fig:s-sbar} shows the
asymmetry $s(x)-\bar{s}(x)$ obtained in these fits.  We note that the
best fit (set $-0$) is quite similar to preliminary results obtained
by the MSTW collaboration \cite{Thorne:2007de}.  It should be
emphasized that a wider range of shapes is obtained if one allows for
a variation of the small-$x$ behavior of $s(x)-\bar{s}(x)$, which is
not well constrained by data.  This is documented in a previous study
by CTEQ \cite{Olness:2003wz} and shown in the right panel of
Fig.~\ref{fig:s-sbar}.

\subsection{Polarized parton densities} 
\label{sec:pol-pdfs}

The polarization of strange quarks and antiquarks in the proton is not
well known at present, for similar reasons as in the case of
unpolarized parton densities.  Many determinations of polarized PDFs
in the literature, such as those in
\cite{Gluck:2000dy,Bluemlein:2002be,Leader:2005ci}, are restricted to
the structure functions for inclusive DIS with electron or muon beams,
which does not permit a separate determination of strange densities.
This is highlighted in the ``valence'' scenario of
\cite{Gluck:2000dy}, which assumes $\Delta\bar{s}(x) = \Delta s(x) =0$
at the starting scale $\mu= 0.63\gev$ of evolution.  This gives a tiny
moment $\mom{\Delta s+\Delta\bar{s}} \approx -4 \times 10^{-3}$ at
$\mu= 1\gev$.
The study in \cite{Hirai:2006sr} additionally fits RHIC data for
$\pi^0$ production, which has no particular sensitivity to strangeness
either.  A process that is specifically sensitive to strangeness is
semi-inclusive kaon production in DIS, which has been measured by
HERMES \cite{Airapetian:2004zf}.  The study by de~Florian et al.\
\cite{deFlorian:2005mw} includes these data and gives two sets of fits
corresponding to different fragmentation functions.  It does not
assume flavor SU(3) symmetry in the polarized sea, allowing
$\Delta\bar{s}(x)$ to differ from $\Delta\bar{u}(x)$ and
$\Delta\bar{d}(x)$.
We note that a recent analysis of semi-inclusive hadron production by
COMPASS reported evidence that the polarized sea is not flavor
symmetric and that $\mom{\Delta\bar{u}}$ and $\mom{\Delta\bar{d}}$ may
have opposite sign \cite{Alekseev:2007vi}.
All analyses performed so far assume $\Delta\bar{s}(x) = \Delta s(x)$.

In Table~\ref{tab:pol-mom} we list the values obtained in recent
analyses for the first moment $\mom{\Delta s+\Delta\bar{s}}$.  Note
that $1/2$ times this moment gives the contribution of strange quarks
and antiquarks to the total spin $1/2$ of the nucleon.  The values
from the analyses
\cite{Gluck:2000dy,Bluemlein:2002be,Leader:2005ci,Hirai:2006sr} have
been obtained
with parameterizations where the polarized sea quark densities
are equal,
$\Delta\bar{u}(x) = \Delta\bar{d}(x) = \Delta\bar{s}(x)$,
so that they should not be regarded as specific determinations of the
polarization of the strange sea.  Rather, they indicate that the
contribution of sea quarks to the nucleon spin is negative and of
moderate magnitude.  The different results of the study
\cite{deFlorian:2005mw} illustrate that a flavor decomposition of this
contribution is currently affected with considerable uncertainties.
The numbers do not suggest that the strangeness contribution to the
nucleon spin is very much suppressed compared with the light flavors
$\bar{u}$ and $\bar{d}$, but further data and analyses are clearly
necessary to settle this issue.  As a further word of caution we
remark that an important fraction of the moments in
Table~\ref{tab:pol-mom} comes from the region of small $x$, where the
polarized densities are not constrained by data.  Quantitative
discussions are given in \cite{Airapetian:2004zf,Bluemlein:2002be}.

\begin{table} 
\vspace{2em}
\renewcommand{\arraystretch}{1.2} 
\begin{center}
\begin{tabular}{|c|ccccccc|} \hline
 & GRSV 2000 \protect\cite{Gluck:2000dy}
 & \multicolumn{2}{c}{BB \protect\cite{Bluemlein:2002be}}
 & \multicolumn{2}{c}{LSS 05 \protect\cite{Leader:2005ci}}
 & \multicolumn{2}{c|}{AAC 06 \protect\cite{Hirai:2006sr}} \\
      set & ``standard'' & 3 & 4 & 1 & 2 & 1 & 2 \\
\hline
$\mom{\Delta s+\Delta\bar{s}} = 2 \mom{\Delta\bar{q}}$
 & $-0.126$ & $-0.148$ & $-0.143$
 & $-0.122$ & $-0.140$ & $-0.10$ & $-0.12$ \\
\hline
\end{tabular} 
\end{center} 
\vspace{0.1em}
\begin{center} 
\parbox{0.55\textwidth}{
  \caption{\label{tab:pol-mom} Lowest moments of polarized parton
    densities.  All analyses shown set $\Delta s(x) =
    \Delta\bar{s}(x)$, and all except for
    \protect\cite{deFlorian:2005mw} take $\Delta\bar{u}(x) =
    \Delta\bar{d}(x) = \Delta\bar{s}(x)$.  The corresponding PDFs have
    been determined at NLO accuracy in the $\overline{\mathrm{MS}}$
    scheme and refer to the scale $\mu= 1\gev$.}  
}
\hspace{1em} 
\begin{tabular}{|c|cc|} \hline
    & \multicolumn{2}{c|}{DNS
      \protect\cite{deFlorian:2005mw,Sassot:2007pc}} \\
set & KRE & KKP \\
\hline
 $\mom{\Delta s+\Delta\bar{s}}$ & $-0.095$ & $-0.090$ \\
 $\mom{\Delta\bar{u}}$          & $-0.046$ & $\phantom{-} 0.076$ \\
 $\mom{\Delta\bar{d}}$          & $-0.048$ & $-0.101$ \\ \hline
\end{tabular} 
\end{center} 
\end{table}


\section{Theoretical approaches} 
\label{sec:theo}

\subsection{Electromagnetic form factors} 
\label{sec:theo-ff}

The strangeness contributions to the electromagnetic and axial form
factors of the nucleon have been studied in a large number of
theoretical approaches (with many studies focusing on the strange
magnetic moment or the electric charge radius) .  Detailed reviews and
discussions can be found in
\cite{Beck:2001dz,Spayde:2003nr,RamseyMusolf:2005rz}.  In
Table~\ref{tab:models} we list a small number of recent results for
the strange Dirac form factor at $-t = 0.1\gev^2$.  We find a
substantial spread between these results and remark that several of
them are outside the range $-0.009 \le F_1^s \le 0.015$ obtained from
the experimental values \eqref{eq:FF-data-1} and \eqref{eq:FF-data-2}.

\begin{table} 
\renewcommand{\arraystretch}{1.2}
\caption{\label{tab:models} Theoretical results for the strange
  Dirac form factor at $t_0= -0.1 \gev^2$.  The numbers for
  Refs.~\protect\cite{Lyubovitskij:2002ng,Silva:2001st,%
    Hammer:1995de,Dubnicka:2006qf,Lewis:2002ix} have been read off
  from graphs.  The value in the last row has been obtained from
  $G_E^s(t_0)$ in \protect\cite{Leinweber:2006ug} and $G_M^s(0)$ in
  \protect\cite{Leinweber:2004tc} using the approximation $G_M^s(t_0)
  \approx G_M^s(0)$, which was also made in Fig.~2 of
  \protect\cite{Young:2007zs}.  Taking into account that $|G_M^s(t_0)|
  < |G_M^s(0)|$ would increase the value of $F_1^s(t_0)$.}
\begin{center} 
\begin{tabular}{|lcc|} \hline
 Approach & reference & $F_1^s\bigl( t= -0.1 \gev^2 \bigr)$  \\
\hline 
Perturbative chiral quark model & \cite{Lyubovitskij:2002ng}
 & $\msp 0.003 \phantom{(0)}$ \\[0.2em]
Chiral quark soliton model ($\pi$) & \cite{Silva:2001st}
 & $\msp 0.063 \phantom{(0)}$ \\[0.2em]
Chiral quark soliton model ($K$)   & \cite{Silva:2001st}
 & $\msp 0.028 \phantom{(0)}$ \\[0.2em]
Vector meson dominance      & \cite{Hammer:1995de}  
 & $-0.07 \phantom{0(0)}$ \\[0.2em]
Vector meson dominance      & \cite{Dubnicka:2006qf}
 & $\msp 0.014 \phantom{(0)}$ \\[0.2em]
Lattice & \cite{Lewis:2002ix}
 & $\msp 0.015(5)$ \\[0.2em]
Lattice $+$ measured magnetic moments and charge radii
 & \cite{Leinweber:2004tc,Leinweber:2006ug}
 & $\msp 0.000(6)$ \\[0.2em]
\hline 
\end{tabular} 
\end{center} 
\end{table} 

The calculation of strange form factors is challenging in many
theoretical approaches.  A large number of studies are based on the
meson cloud picture, where the nucleon fluctuates into a $K$ and a
$\Sigma$ or $\Lambda$.  The coupling to the strangeness current then
proceeds through valence degrees of freedom, namely the $\bar{s}$ in
the kaon and the $s$ in the hyperon.  Concerns have been raised about
the quantitative reliability of such calculations, based on the
possible importance of unitarity corrections \cite{Musolf:1996qt} and
of higher-mass states \cite{Geiger:1996re,Barz:1998ih} such as $K^*$
mesons \cite{Barz:1998ih}.  There seems to be no consensus about these
issues in the literature, see \cite{Melnitchouk:1999mv,Forkel:1999kz}
and \cite{RamseyMusolf:2005rz}.  We will not quantitatively use the
meson cloud picture in the present work, but use it as a qualitative
guide in Sect.~\ref{sec:mod-ans}.  Chiral perturbation theory provides
a systematic framework for calculations in terms of hadron degrees of
freedom, but its predictive power for strange form factors is limited,
as discussed in \cite{Kubis:2005cy}.

Kaons also play an essential role in chiral quark models such as the
one in \cite{Lyubovitskij:2002ng}, where the nucleon is described in
terms of three constituent quarks coupling to the pseudo-Goldstone
bosons.  In this approach, nonzero strange form factors are due to the
splitting of a $u$ or $d$ quark into a kaon and an $s$ quark.  The
chiral quark soliton model \cite{Silva:2001st} does not rely on the
constituent quark picture, containing as degrees of freedom both
quarks and antiquarks coupling to pions and kaons.

A different approach is based on dispersion relations, which represent
the form factors for spacelike $t$ in terms of an integral over their
imaginary parts in the timelike region.  The assumption that the
dispersion integral is dominated by single vector meson states leads
to the vector meson dominance approximation, which underlies many
calculations of the strange form factors.  A typical procedure is to
fix the relevant nucleon-meson coupling constants from the isoscalar
electromagnetic form factors of the nucleon and then to predict the
form factors of the strangeness current.  Such analyses often obtain
rather large couplings of the nucleon to the $\omega(782)$ and the
$\phi(1020)$, see for instance \cite{Hammer:1995de}.  These large
couplings are in strong conflict with determinations from
nucleon-nucleon potential studies \cite{Nagels:1978me,Dumbrais:1983jd}
or from dispersion relations for forward nucleon-nucleon scattering
\cite{Grein:1979nw}, with SU(6) symmetry \cite{Nagels:1973rq}, and in
the case of the $\phi$ with the OZI rule.
To understand why fits of nucleon form factors with a small number of
vector meson resonances can give large couplings, we consider the
simplified case of just two mesons with masses $m_1$ and $m_2$,
\begin{align}
\frac{a_1}{m_1^2 - t} + \frac{a_2}{m_2^2 - t}
 &= \frac{a_1 (m_2^2 - m_1^2) - (a_1+a_2) (t - m_1^2)}{%
          m_1^2\ms m_2^2 - (m_1^2+m_2^2)\, t + t^2} \,.
\end{align}
In order to obtain a $1/t^2$ behavior of the isoscalar form factor at
large $-t$, one must have $a_1\approx -a_2$ to keep the term with $t$
in the numerator small.  At $t=0$ the form factor is then
approximately given by $a_1 (m_2^2 - m_1^2) /(m_1^2\ms m_2^2)$, and the
small mass difference between $\phi(1020)$ and $\omega(782)$ forces
the couplings $a_1$ and $a_2$ to be large.  Taking into account
higher-mass resonances significantly reduces this trend.  As an
illustration, we have fitted the isoscalar nucleon form factors to a
sum of contributions from $\phi(1020)$, $\omega(782)$ and
$\omega(1420)$, with or without an additional contribution from
$\omega(1650)$.
We require an asymptotic behavior $F_1^p+F_1^n \sim 1/t^2$ at large
 $-t$, which provides a linear relation between the different meson
 couplings in generalization of the simple case we just discussed.
With the normalization condition $F_1^p(0)+F_1^n(0) = 1$ this leaves
two free parameters if the $\omega(1650)$ is included and a single
parameter if this resonance is omitted.  For the meson-nucleon
couplings relevant to the Dirac form factors we obtain
\begin{align}
g_{\phi NN}^V &= -9.13 \,, &  g_{\omega NN}^V &= 20.6
\intertext{in the fit without $\omega(1650)$, where the couplings
  refer to the ground state mesons and are denoted by
  $\smash{G^V_{NN\phi}}$ and $G^V_{NN\omega}$ in
  \protect\cite{Dumbrais:1983jd}.  Including the $\omega(1650)$, we
  obtain a good description of the data with}
g_{\phi NN}^V &=  4.69 \,, &  g_{\omega NN}^V &= 11.9 \,,
\end{align}
where $g_{\omega NN}^V$ is fixed to a value as small as the data
permits, in order to minimize the tension with the still lower values
obtained in \cite{Nagels:1978me,Grein:1979nw,Nagels:1973rq}.  Taking
both couplings as free fit parameters we find $g_{\phi NN}^V = -0.06$
and $g_{\omega NN}^V = 14.9$.  Similar values have been obtained in
\cite{Dubnicka:2002yp}.

This simple exercise
suggests to take with great care the corresponding predictions for
strange form factors, where contributions from $\phi$ resonances are
strongly enhanced compared with those from $\omega$ states.  A more
realistic treatment requires the inclusion of continuum states, as has
for instance been done in \cite{Hammer:1999uf,Belushkin:2006qa}.  The
analysis in \cite{Belushkin:2006qa} found that the inclusion of
the $K\bar{K}$ and $\rho\pi$ continua makes the interpretation of a
residual $\phi$ resonance contribution ambiguous.


\subsection{The strangeness asymmetry $s(x) - \bar{s}(x)$}
\label{sec:s-sbar-theo}

The meson cloud picture naturally induces an asymmetry in the momentum
distribution of strange quarks and antiquarks, which was first
observed in \cite{Signal:1987gz} and underlies many calculations
\cite{Holtmann:1996be,Christiansen:1998dz,Avila:2007mf,Brodsky:1996hc,%
  Ding:2004ht,Cao:1999da,Cao:2003ny}.  In this picture, the densities
of $s$ and 
$\bar{s}$ in the nucleon are given as convolutions of the longitudinal
momentum distributions of the kaon or hyperon within the nucleon and
the valence distribution of $\bar{s}$ in the kaon or of $s$ in the
hyperon.  A similar mechanism is realized in chiral quark models
\cite{Ding:2004dv},
where the constituent quarks of the proton can fluctuate into a kaon
and a strange quark.  There is a significant spread among meson cloud
model predictions for the shape of $s(x)-\bar{s}(x)$, including its
sign and the number of zero crossings, see e.g.\ the comparative study
in \cite{Cao:1999da}.  The inclusion of $K^*$ fluctuations in
\cite{Cao:2003ny} also had a significant effect, changing even the
sign of the momentum asymmetry $\mom{x (s-\bar{s})}$ compared to the
result with kaon fluctuations alone.  We note that the predictions for
$s-\bar{s}$ in such models typically have a zero at a value of $x$
much larger than $0.01$ and are thus rather different from the results
obtained in the PDF fits \cite{Lai:2007dq,Thorne:2007de}.

The study \cite{Catani:2004nc} pointed out that in perturbative
evolution at three-loop accuracy and beyond, graphs with three-gluon
exchange in the $t$-channel generate an $s-\bar{s}$ asymmetry.
Starting with $s(x) = \bar{s}(x)$ at the low scale $\mu= 0.51 \gev$,
the authors of \cite{Catani:2004nc} find that $s(x)-\bar{s}(x)$ for
$\mu\ge 2\gev$ is positive at small and negative at intermediate to
large $x$, with $\langle x (s-\bar{s}) \rangle \approx -5 \times
10^{-4}$.  This is much smaller than the central fit results in
\cite{Lai:2007dq,Thorne:2007de}, which suggests that this perturbative
mechanism plays only a minor role in the generation of the momentum
asymmetry.


\section{Relating the strange Dirac form factor to $s(x)-\bar{s}(x)$} 
\label{sec:our-model}

We now formulate a model ansatz for the $C$ odd part of the
generalized parton distribution $H^s$ at zero skewness, which will
allow us to calculate the Dirac form factor $F_1^s$ from the
phenomenologically extracted asymmetry $s-\bar{s}$ of momentum
distributions.  Following previous studies of generalized parton
distributions in the non-strange sector
\cite{Goeke:2001tz,Burkardt:2002hr,Diehl:2004cx,Guidal:2004nd}, we
assume an exponential $t$ dependence with an $x$ dependent slope and
set
\begin{equation}
  \label{master-ansatz}
H^s(x,t) - H^{\bar{s}}(x,t) =
  \bigl[ s(x) - \bar{s}(x) \bigr]\, e^{\ms t f_s(x)} \,,
\end{equation}
where for the slope we take the simple form
\begin{equation}
  \label{simple-prof}
f_s(x) = \alpha' (1-x) \log\frac{1}{x} \,,
\end{equation}
which was already proposed in \cite{Burkardt:2002hr}.  With
\eqref{impact-def} it is easy to see that the profile function
$f_s(x)$ has a simple physical interpretation in terms of the average
squared impact parameter
\begin{align}
\langle b^2 \rangle_x
&= \frac{\int d^2\boldsymbol{b}\; \boldsymbol{b}^2
  \bigl[ s(x,b) - \bar{s}(x,b) \bigr]}{\int d^2\boldsymbol{b}\,
  \bigl[ s(x,b) - \bar{s}(x,b) \bigr]}
 = 4 f_s(x)
\end{align}
associated with the difference between $s$ and $\bar{s}$
distributions.  As shown in \cite{Burkardt:2004bv}, a finite average
transverse size of parton configurations with $x\to 1$ in the nucleon
requires $\langle b^2 \rangle_x$ to vanish at least like $(1-x)^2$ in
this limit, which is obviously satisfied for the ansatz
\eqref{simple-prof}.

In the opposite limit $x\to 0$, we use simple Regge phenomenology as a
guide for our parameterization.  The form \eqref{simple-prof}
corresponds to the behavior $H^s - H^{\bar{s}} \sim x^{-\alpha(t)}$,
which arises from the exchange of a single Regge pole with a linear
trajectory $\alpha(t) = \alpha(0) + \alpha' t$, or from the
superposition of several Regge poles with the same value of $\alpha'$.
This is a generalization to finite $t$ of a small-$x$ behavior
$x^{-\alpha}$ for the usual parton densities, which is consistent with
phenomenology.  The leading Regge trajectory that can contribute to
$H^s - H^{\bar{s}}$ is the one for the $\phi$ mesons, and assuming a
linear form $\alpha_\phi(t) = \alpha_\phi(0) + \alpha' t$ one obtains
$\alpha_\phi(0) = 0.13$ and $\alpha' = 0.84 \gev^{-2}$ from the masses
and spins of $\phi(1020)$ and $\phi_3(1850)$.  This value of $\alpha'$
is close to the one for other meson trajectories, such as the ones for
the $\rho$ and $\omega$.  The $\phi$ trajectory contributes to soft
hadronic scattering processes like kaon-nucleon scattering or
photoproduction of the $\phi$ meson.  It is however neglected in most
analyses of these processes (which is justified if the
$\phi\ms$-nucleon coupling is sufficiently small, see our discussion
in Sect.~\ref{sec:theo-ff}).  An exception is the analysis of the
total kaon-nucleon cross sections performed by Barger and Olsson
\cite{Barger:1966nn}, who found an intercept $\alpha_\phi(0) = 0.33
\pm 0.06$ of similar size as the result obtained from the hadronic
spectrum.  We emphasize that in our approach we do not need an
explicit value for the $\phi\ms$-nucleon coupling, since the
normalization in \eqref{master-ansatz} is fixed by the difference
$s-\bar{s}$ of parton distributions.

We note that the CTEQ6.5S densities at $\mu=2\gev$ are well
approximated by
\begin{equation}
  \label{s-sbar_small-x}
x (s-\bar{s}) \approx a x^{0.28}
\end{equation}
in the region $10^{-5} < x < 10^{-4}$, with $a = -0.031, +0.023,
-0.044$ for the respective sets $-0,-1,-2$.  This corresponds to
$\alpha(0) = 0.72$, which is quite far from the values we estimate for
the $\phi$ trajectory.  There are however no experimental constraints
for the behavior of $s-\bar{s}$ at very small $x$, and a value of
$\alpha(0)$ between $0.1$ and $0.4$ is within the range for which a
good description of all relevant data has been obtained in the CTEQ
study \cite{Olness:2003wz}.

Using an ansatz as in \eqref{master-ansatz} for the valence
combinations $H^u-H^{\bar{u}}$ and $H^d-H^{\bar{d}}$, we obtained in
\cite{Diehl:2004cx} a good description of the electromagnetic Dirac
form factors of proton and neutron.  Given the wealth of data in this
case, we chose in that study more complicated forms than
\eqref{simple-prof} for the profile functions $f_u(x)$ and $f_d(x)$.
We find that for $10^{-4}<x<0.1$ they are both very well approximated
by the form \eqref{simple-prof} with $\alpha'= 1\gev^{-2}$, which
remains close to $f_u(x)$ for $x>0.1$.  Given the fast decrease of
$s(x) - \bar{s}(x)$ with $x$, the small-$x$ region turns out to be
most important for our calculation of $F_1^s$.

\begin{figure}
\begin{center}
  \psfrag{xx}[c][c]{\small $-t\; [\gev^2]$}
  \psfrag{tt}[c][c]{\small $F_1^s(t)$}
  \psfrag{bb}[r][c]{\footnotesize $-0.5\, F_1^n(t)$}
\includegraphics[width=0.49\textwidth,%
  bb=80 50 385 295]{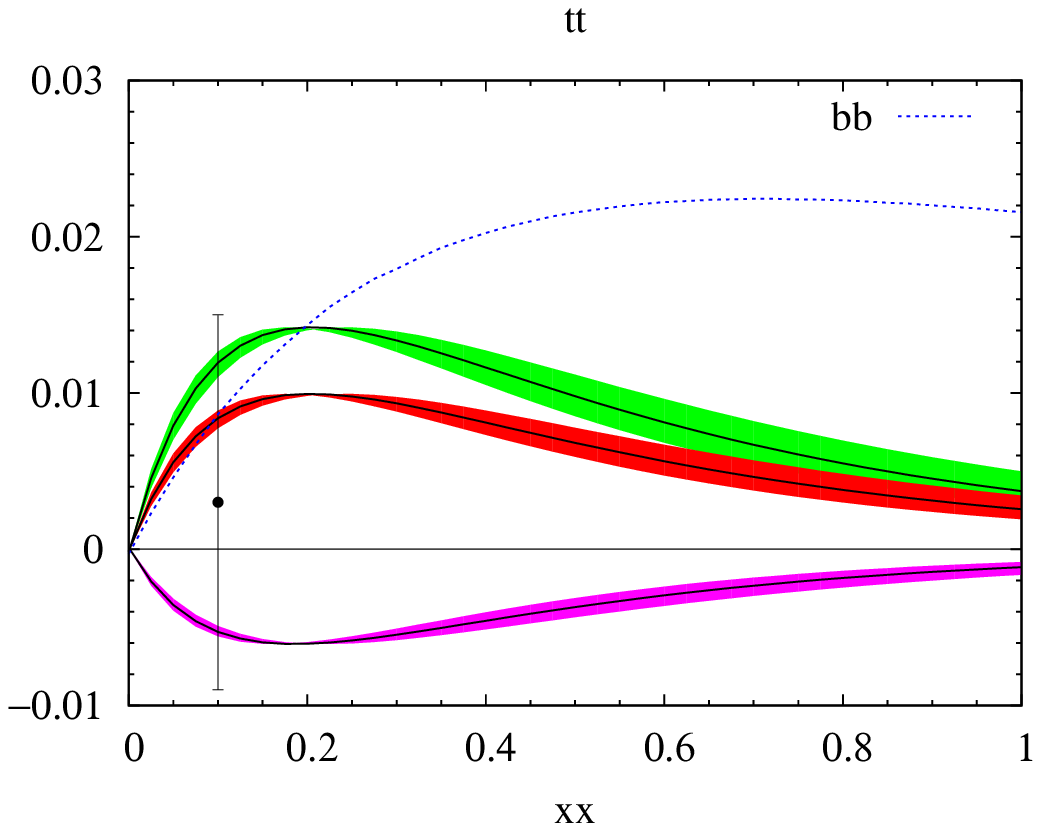}
\includegraphics[width=0.49\textwidth,%
  bb=80 50 385 295]{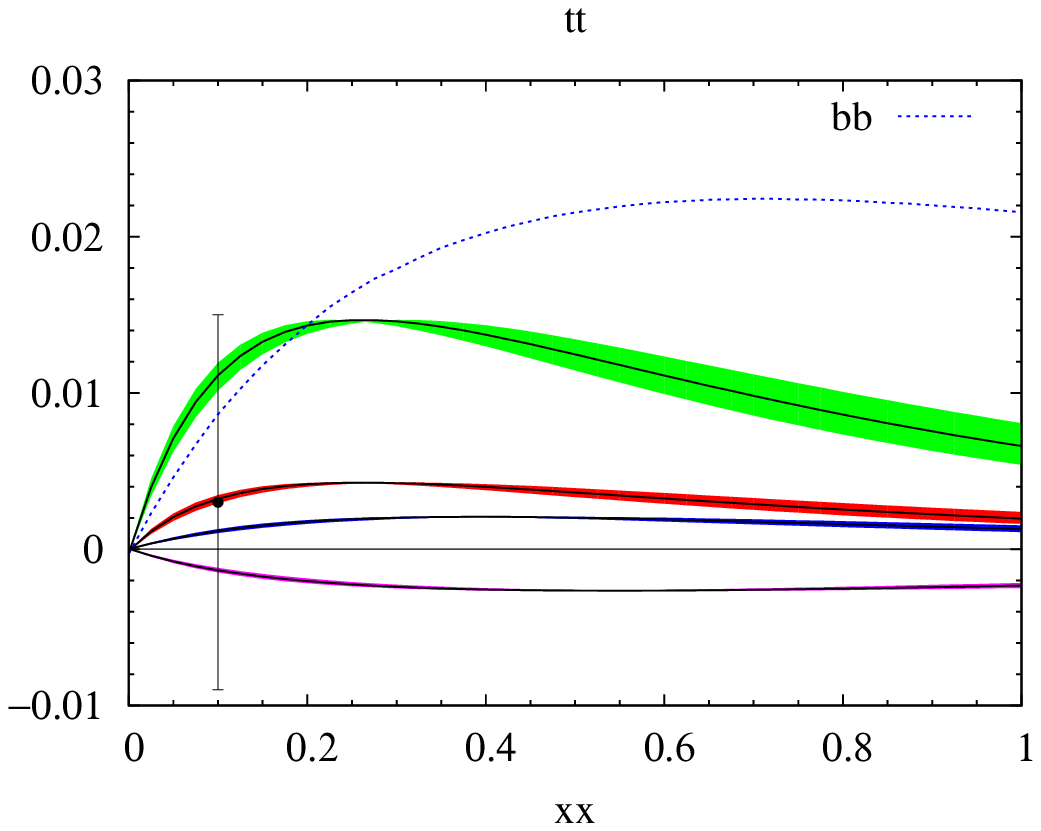}
\end{center}
\caption{\label{fig:F1s-default} The strange Dirac form factor
  obtained from the sum rule \protect\eqref{Dirac} with the model
  ansatz in \protect\eqref{master-ansatz} and
  \protect\eqref{simple-prof}.  Central curves are for $\alpha'=
  1\gev^{-2}$ and bands for $0.85 \gev^{-2} < \alpha' < 1.15
  \gev^{-2}$.  The corresponding parton densities are (from top to
  bottom): CTEQ6.5S sets $-2$, $-0$, $-1$ in the left panel and CTEQ6
  sets B$+$, B, C, B$-$ in the right panel.  The data point is our
  estimate \protect\eqref{eq:FF-data-1} of $F_1^s$, and the dotted
  curve corresponds to the parameterization of $F_1^n$ discussed in
  the text.}
\end{figure}

We take the ansatz \eqref{master-ansatz} with the CTEQ6.5S densities
\cite{Lai:2007dq} at $\mu= 2\gev$ as input, where the chosen scale is
to be considered as a compromise between a small value appropriate for
arguments based on hadronic Regge phenomenology and a large value
where the densities are sufficiently constrained by experimental data.
In the left panel of Fig.~\ref{fig:F1s-default} we show the values of
$F_1^s(t)$ obtained with the best fit (set $-0$) and with the
alternative fits (sets $-1$ and $-2$).  The central curves are for
$\alpha'=1 \gev^{-2}$ in \eqref{simple-prof}, and the bands correspond
to $\alpha'$ between $0.85 \gev^{-2}$ and $1.15 \gev^{-2}$.  We regard
this variation as an estimate of the parametric uncertainty within our
model, with the lower value corresponding to the estimate of the
$\phi$ trajectory from the meson masses.  In the following we refer to
the result with $\alpha'=1 \gev^{-2}$ and CTEQ6.5S set $-0$ as our
default prediction.

If instead of taking \eqref{simple-prof} we set $f_s(x)$ equal to the
profile functions $f_u(x)$ or $f_d(s)$ obtained in
\cite{Diehl:2004cx}, the form factor lies within the bands in the
figure, except in the region where $|F_1^s(t)|$ has its maximum.  In
that region, the difference of $F_1^s(t)$ obtained with the different
profile functions just mentioned is at most $5\%$.
As a further alternative, we have made the ansatz
\eqref{master-ansatz} at scale $\mu= 1.3\gev$, which is the starting
scale of the CTEQ parameterizations.  Taking the profile function
\eqref{simple-prof} with $\alpha'= 1\gev^{-2}$, we again obtain values
within the bands of Fig.~\ref{fig:F1s-default}, except for deviations
of up to 5\% around the maximum of $|F_1^s(t)|$.  Clearly, the largest
spread in predictions for $F_1^s(t)$ within our model is due to the
different parton densities used as input.  To further explore this, we
have taken the parameterizations from the CTEQ6 study
\cite{Olness:2003wz} at $\mu= 2 \gev$, which provides a wider range of
shapes as we have seen in Fig.~\ref{fig:s-sbar}.  The resulting curves
for $F_1^s(t)$ are shown in the right panel of
Fig.~\ref{fig:F1s-default}.  We recall that sets $-1$ and $-2$ in
\cite{Lai:2007dq} and sets B$-$ and B$+$ in \cite{Olness:2003wz} were
chosen to minimize or maximize the moment $\mom{x (s-\bar{s})}$.  They
are hence not preferred, although consistent with the data fitted in
\cite{Lai:2007dq,Olness:2003wz}.

\begin{figure}
\begin{center}
  \psfrag{xx}[c][c]{\small $-t\; [\gev^2]$}
  \psfrag{tt}[c][c]{}
  \psfrag{bb}[r][c]{\footnotesize $F_1^n(t)$}
  \psfrag{dd}[r][c]{\footnotesize $-4.5\, F_1^s(3.5\, t)$}
\includegraphics[width=0.49\textwidth,%
  bb=80 50 385 295]{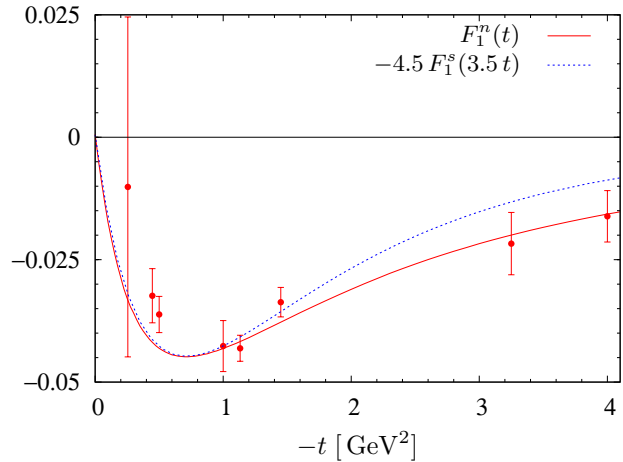}
\end{center}
\caption{\label{fig:F1n} Data for the neutron Dirac form factor and
  fit from \protect\cite{Diehl:2004cx}.  The dotted line corresponds
  to our default prediction for $F_1^s$, obtained with $\alpha'=
  1\gev^2$ and CTEQ6.5S set $-0$.  The curve is rescaled such that its
  minimum coincides with the one of $F_1^n$.}
\end{figure}

We see that in all cases the form factor $F_1^s(t)$ is quite small and
fully compatible with the estimates \eqref{eq:FF-data-1} and
\eqref{eq:FF-data-2} extracted from experiment.  We remark that for
most of our curves, a linear behavior $F_1^s(t) \propto t$ as in
\eqref{eq:FF-data-2} is not a good approximation for $-t$ much above
$0.1 \gev^2$.

It is instructive to compare our results with another small nucleon
form factor, namely the Dirac form factor $F_1^n(t)$ of the neutron.
Figure~\ref{fig:F1n} shows data together with the default fit from
\cite{Diehl:2004cx}, which we already mentioned in connection with the
profile functions $f_u(x)$ and $f_d(x)$.  The same parameterization
multiplied by $-0.5$ is shown as a dotted curve in
Fig.~\ref{fig:F1s-default}.  We recall at this point that
\begin{equation}
F_1^n(t) = \tfrac{2}{3} F_1^d(t) - \tfrac{1}{3} F_1^u(t)
         - \tfrac{1}{3} F_1^s(t) \,,
\end{equation}
where the labels on the r.h.s.\ indicate the quark flavor
contributions to the Dirac form factor $F_1^p$ of the \emph{proton}.
The fit in \cite{Diehl:2004cx} neglected the strange contribution in
$F_1^n$ and in $F_1^p$.  We see in Fig.~\ref{fig:F1s-default} that our
estimates for $F_1^s$ are at most half as large in magnitude as
$F_1^n$ for $-t= 0.2\gev^2$, so that at that point $-\frac{1}{3} F_1^s$
contributes at most $1/6$ to the neutron form factor.  For higher $-t$
we find that $F_1^s(t)$ decreases faster than $F_1^n(t)$, which we
will explain shortly.  Only at small $t$ do we find a stronger
influence of the strangeness contribution.  If our estimate is
correct, this is of relevance for the flavor analysis of the Dirac
radius of the neutron, which in more familiar terms can be expressed
through the electric radius and a contribution from the magnetic
moment,
\begin{align}
\langle r^2 \rangle_{1}^{n}
&\;=\; 6\, \frac{d}{dt} F_1^n(t) \biggl|_{t=0}
 \;=\; 6\, \frac{d}{dt} G_E^n(t) \biggl|_{t=0}
         - \frac{3 \mu_n}{2 m_n^2}
 \;\approx\; \langle r^2 \rangle_{E}^{n} + 0.127 \fm^{2} \,.
\end{align}
Concerning the proton form factor, we find that the values of $F_1^s(t)$
shown in Fig.~\ref{fig:F1s-default} amount to at most $3\%$ of
$F_1^p(t)$ at any $t$.


\subsection{The shape of $F_1^s(t)$}

Let us now discuss the general features of $F_1^s(t)$ that emerge with
our model ansatz, where we have
\begin{align}
  \label{F1s-int}
F_1^s(t) &= \int_0^1 dx\,
  \bigl[ s(x) - \bar{s}(x) \bigr] \, e^{\ms t f_s(x)} \,,
\\
  \label{F1s-der-int}
\frac{d}{dt}\ms F_1^s(t) &= \int_0^1 dx\,
  \bigl[ s(x) - \bar{s}(x) \bigr] \ms f_s(x)\, e^{\ms t f_s(x)} \,.
\end{align}
For the neutron form factor the situation is slightly more complicated
even if we neglect the strangeness contribution, since the fit in
\cite{Diehl:2004cx} required different profile functions for $u$ and
$d$ quarks.  Since their difference is only moderate, the discussion
of $F_1^n(t)$ is however quite similar.

With $f_s(x)$ being a decreasing function, the factor $e^{\ms t
  f_s(x)}$ increasingly suppresses small $x$ values in the integral
\eqref{F1s-int} when $-t$ becomes larger.  For increasing $-t$ the
form factor $F_1^s$ is therefore connected with $s-\bar{s}$ at
increasing values of $x$.  With the profile function
\eqref{simple-prof} we obtain the Drell-Yan relation $p =
\half (1+\beta)$ between the powers describing the asymptotic power
laws $F_1^s \sim (-t)^{-p}$ for $t\to -\infty$ and $s-\bar{s} \sim
(1-x)^\beta$ for $x\to 1$ \cite{Burkardt:2004bv,Diehl:2004cx}.  The
difference $s-\bar{s}$ of sea quark distributions falls off faster
with $x$ than the valence distributions $u-\bar{u}$ and $d-\bar{d}$,
which are relevant for $F_1^n$, so that one generally expects
$|F_1^s(t)|$ to decrease faster than $|F_1^n(t)|$ with $-t$.  As
Figs.~\ref{fig:F1s-default} and \ref{fig:F1n} show, this is indeed the
case in our model.

\begin{figure}
\begin{center}
  \psfrag{xx}[c][c]{\small $x$}
  \psfrag{tt}[c][c]{\small $x\ms (H^{s} - H^{\bar{s}})$}
  \psfrag{bb}[l][c]{\footnotesize $t= 0$}
  \psfrag{dd}[l][c]{\footnotesize $\phantom{t}= -0.2 \gev^2$}
  \psfrag{hh}[l][c]{\footnotesize $\phantom{t}= -0.5 \gev^2$}
\includegraphics[width=0.48\textwidth,%
  bb=80 50 385 295]{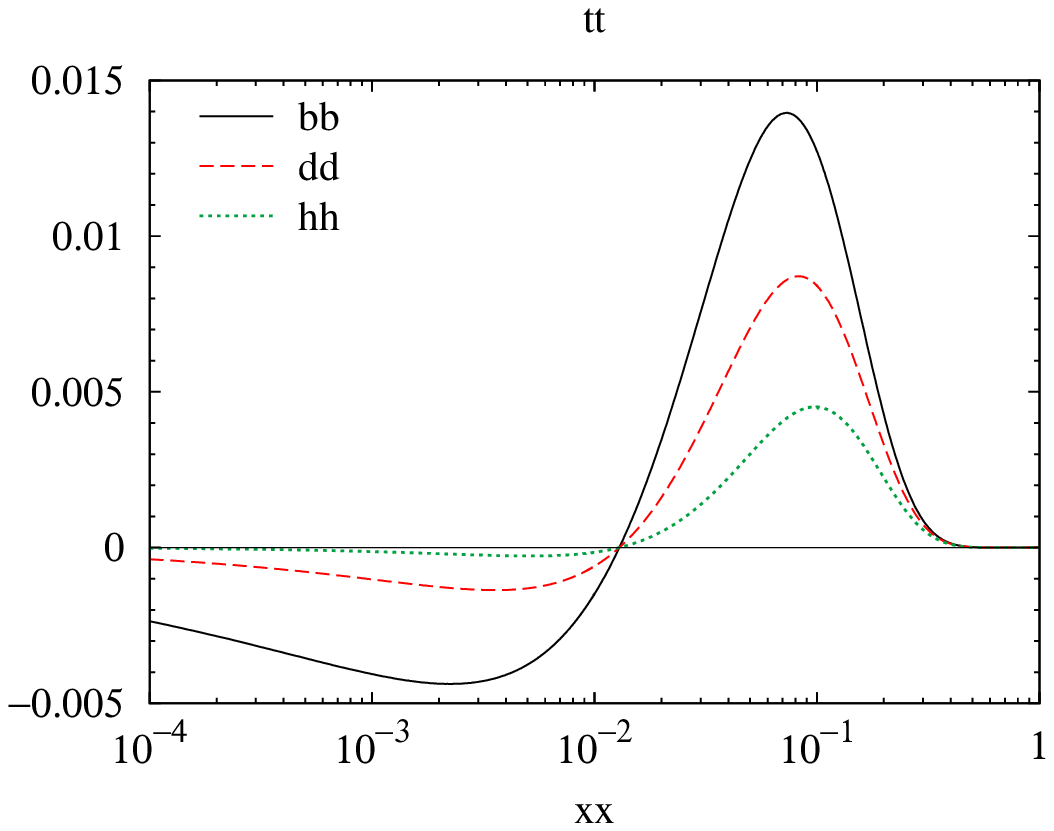}
  \psfrag{tt}[c][c]{\small $x\ms \frac{d}{dt} (H^{s} - H^{\bar{s}})$}
\hspace{0.5em}
\includegraphics[width=0.48\textwidth,%
  bb=80 50 385 295]{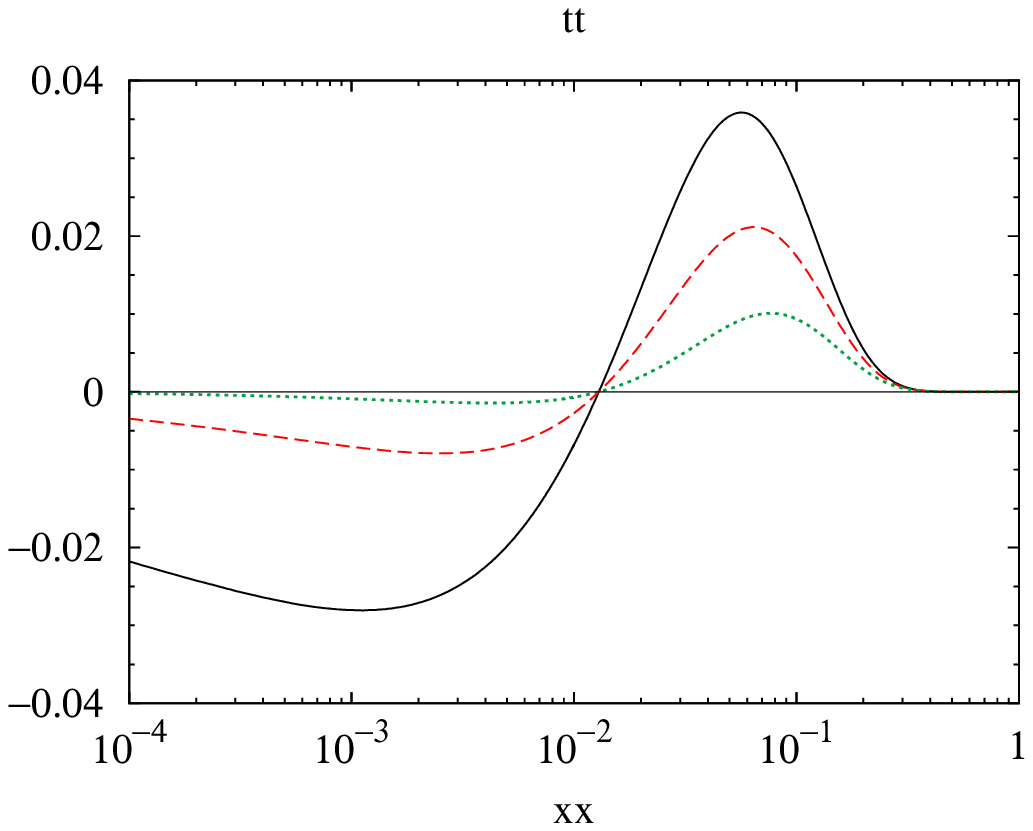}
\\[1em]
  \psfrag{tt}[c][c]{\small
    $\frac{2}{3} x\ms (H^{d} - H^{\bar{d}}\ms)
    -\frac{1}{3} x\ms (H^{u} - H^{\bar{u}})$}
  \psfrag{dd}[l][c]{\footnotesize $\phantom{t}= -0.7 \gev^2$}
  \psfrag{hh}[l][c]{\footnotesize $\phantom{t}= -1.4 \gev^2$}
\includegraphics[width=0.48\textwidth,%
  bb=80 50 385 295]{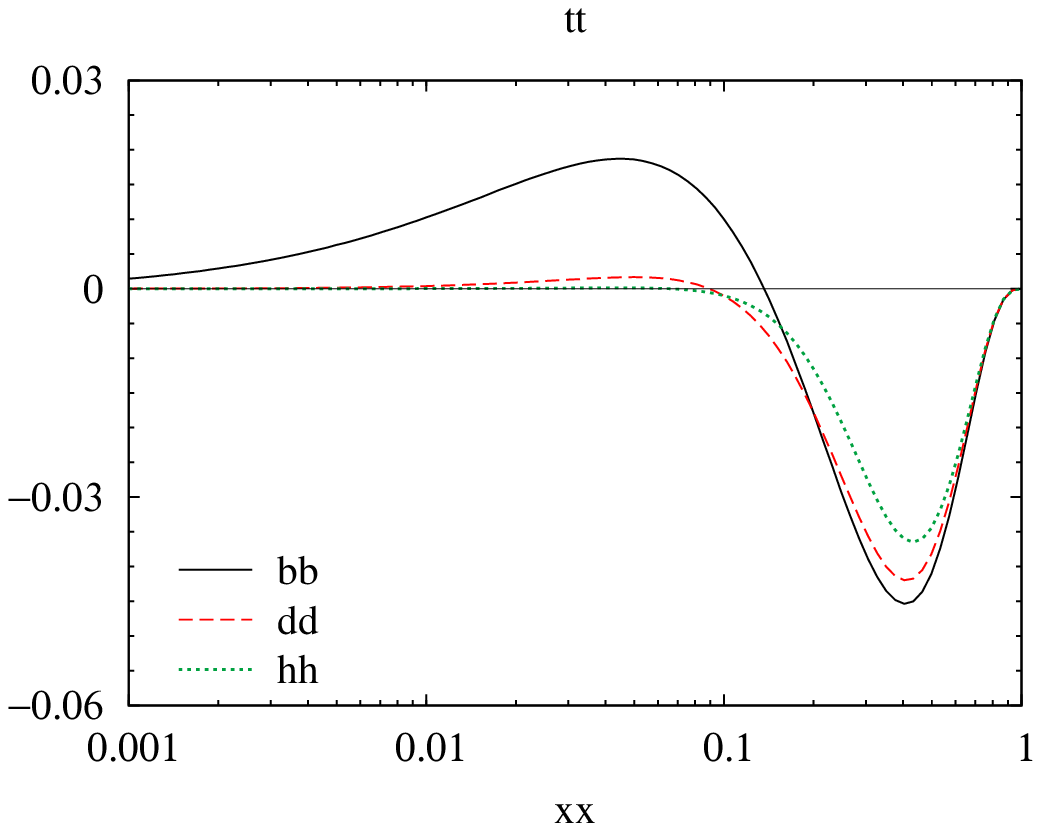}
  \psfrag{tt}[c][c]{\small
    $\frac{2}{3} x\ms \frac{d}{dt} (H^{d} - H^{\bar{d}}\ms )
    -\frac{1}{3} x\ms \frac{d}{dt} (H^{u} - H^{\bar{u}})$}
\hspace{0.5em}
\includegraphics[width=0.48\textwidth,%
  bb=80 50 385 295]{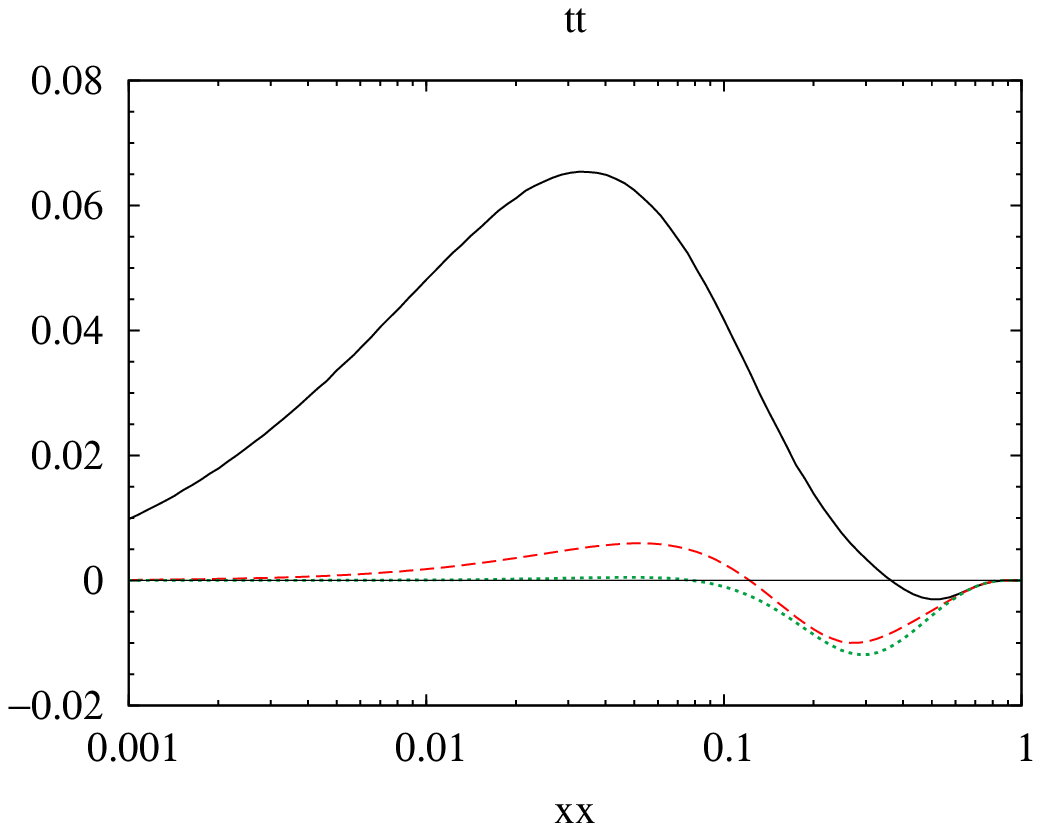}
\end{center}
\caption{\label{fig:integrands} Upper plots: the scaled integrands of
  $F_1^s(t)$ and $d F_1^s(t)/ dt$ in \protect\eqref{F1s-int} and
  \protect\eqref{F1s-der-int}, obtained with our default prediction.
  Lower plots: the corresponding scaled integrands for $F_1^n(t)$ and
  $d F_1^n(t)/ dt$, obtained with the default fit in
  \protect\cite{Diehl:2004cx}.  The values of $t$ in the right panels
  are as for the corresponding curves on the left.}
\end{figure}

In Fig.~\ref{fig:integrands} we show the integrands of $F_1^s$ and $d
F_1^s /dt$ in \eqref{F1s-int} and \eqref{F1s-der-int}.  The integrands
are multiplied with $x$ in the plots, so that with the logarithmic
scale for $x$ we obtain the form factor or its derivative as the area
under the corresponding curve.  For $t=0$ the integrand of $F_1^s$ is
$s(x)-\bar{s}(x)$, which gives a zero integral because of quantum
number constraints.  The integrand for the derivative $d F_1^s /dt$
has an extra factor $f_s(x)$, which enhances small $x$ values relative
to larger ones.  At $t=0$ one therefore has $d F_1^s /dt < 0$ at $t=0$
if $s-\bar{s}$ is negative at small $x$ and positive at large $x$.
This is the case for the CTEQ fits \cite{Lai:2007dq,Olness:2003wz}
except for sets $-1$ and B$-$.  As $-t$ increases, the factor $e^{\ms
  t f_s(x)}$ suppresses small $x$ values in \eqref{F1s-der-int}, and
for sufficiently large $-t$ the derivative $d F_1^s /dt$ has the
opposite sign than at $t=0$.  For some $t$ one hence obtains a maximum
or minimum of $F_1^s(t)$.  The value of $-t$ where this happens is
larger for parameterizations of $s-\bar{s}$ which have the zero
crossing at larger $x$, as one can check by comparing
Figs.~\ref{fig:s-sbar} and \ref{fig:F1s-default}.

Figure~\ref{fig:integrands} also shows the integrands for $F_1^n$ and
$d F_1^n/ dt$ for the default fit in \cite{Diehl:2004cx}.  The
discussion for the sign of the derivative $d F_1^n/ dt$ and the
presence of a minimum of $F_1^n(t)$ proceeds in analogy to the case of
the strangeness form factor.  Since the zero crossing of $\frac{2}{3}
(d-\bar{d}) - \frac{1}{3} (u-\bar{u})$ occurs at much larger $x$ than
the one of $s-\bar{s}$ in the CTEQ6.5S parameterizations, whereas the
respective profile functions are similar, $|F_1^n(t)|$ assumes its
maximum at significantly larger $-t$ than $|F_1^s(t)|$.

Let us at this point mention the PDFs extracted in
\cite{Barone:1999yv}.  In contrast to the analyses by CTEQ
\cite{Lai:2007dq,Olness:2003wz} and MSTW \cite{Thorne:2007de}, the
strangeness asymmetry $x (s-\bar{s})$ in \cite{Barone:1999yv} has a
zero at $0.4 \lsim x \lsim 0.5$ and a maximum at $x\sim 0.7$.  Taking
this distribution with the same profile functions $f_s(x)$ explored
above, we obtain a form factor $F_1^s(t)$ with a very flat maximum
$F_1^s \sim 0.0025$ for $1\gev^{2} \lsim -t\; \lsim\; 2\gev^{2}$ and a
slow decrease with $-t$.  In this case, the bulk of the form factor
integral \eqref{F1s-int} comes from very large $x$, where we deem our
model for the profile function $f_s(x)$ associated with sea quarks
very insecure.  Since the study \cite{Barone:1999yv} used inclusive
cross sections for $\nu$ and $\bar{\nu}$ DIS but no dimuon production
data to constrain $s-\bar{s}$, we do not regard such a scenario as
strongly motivated.  This example illustrates however that within our
model framework, strong changes in $s-\bar{s}$ result in qualitatively
different forms of $F_1^s(t)$, which may eventually be ruled out by
data.

The relations between the $x$ dependence of $s-\bar{s}$ and the $t$
dependence of $F_1^s(t)$ discussed in this subsection follow from the
general features of our ansatz in \eqref{master-ansatz} and
\eqref{simple-prof} and will also hold for more complicated forms.
The neutron form factor $F_1^n(t)$ and the combination $\frac{2}{3}
(d-\bar{d}) - \frac{1}{3} (u-\bar{u})$ of valence distributions, which
are both much better known than their strangeness counterparts,
provide an example where these relations are indeed seen and
corroborate our prediction for the behavior of $F_1^s(t)$ with a given
form of $s(x)-\bar{s}(x)$.


\subsection{A modified ansatz}
\label{sec:mod-ans}

Our ansatz in \eqref{master-ansatz} is special in that it assumes a
$t$ dependence in the form of a global factor multiplying $s(x) -
\bar{s}(x)$.  It implies that the difference $s(x,b) - \bar{s}(x,b)$
of impact parameter densities has a Gaussian shape in $b$ and in
particular does not change sign for given $x$.  One may wonder whether
this ansatz is too restrictive.  The physical picture of meson cloud
models for instance suggests that the typical transverse position of
$s$ is smaller than for $\bar{s}$, since the $\bar{s}$ originates from
a kaon, which due to its smaller mass tends to be at larger distances
than the hyperon containing the $s$.  If this effect is strong
enough, one may have a node of $s(x,b) - \bar{s}(x,b)$ in $b$.

\begin{figure}
\begin{center}
  \psfrag{xx}[c][c]{\small $b\, [\fm]$}
  \psfrag{tt}[c][c]{\small
    $2\pi b \bigl[ s(x,b) - \bar{s}(x,b) \bigr]$ ~in~fm$^{-1}$}
  \psfrag{bb}[r][c]{\footnotesize $A=0\quad           \bar{A}=4$}
  \psfrag{dd}[r][c]{\footnotesize $\phantom{A}=0\quad \phantom{A}=0$}
  \psfrag{hh}[r][c]{\footnotesize $\phantom{A}=4\quad \phantom{A}=0$}
\includegraphics[width=0.49\textwidth,%
  bb=80 50 385 295]{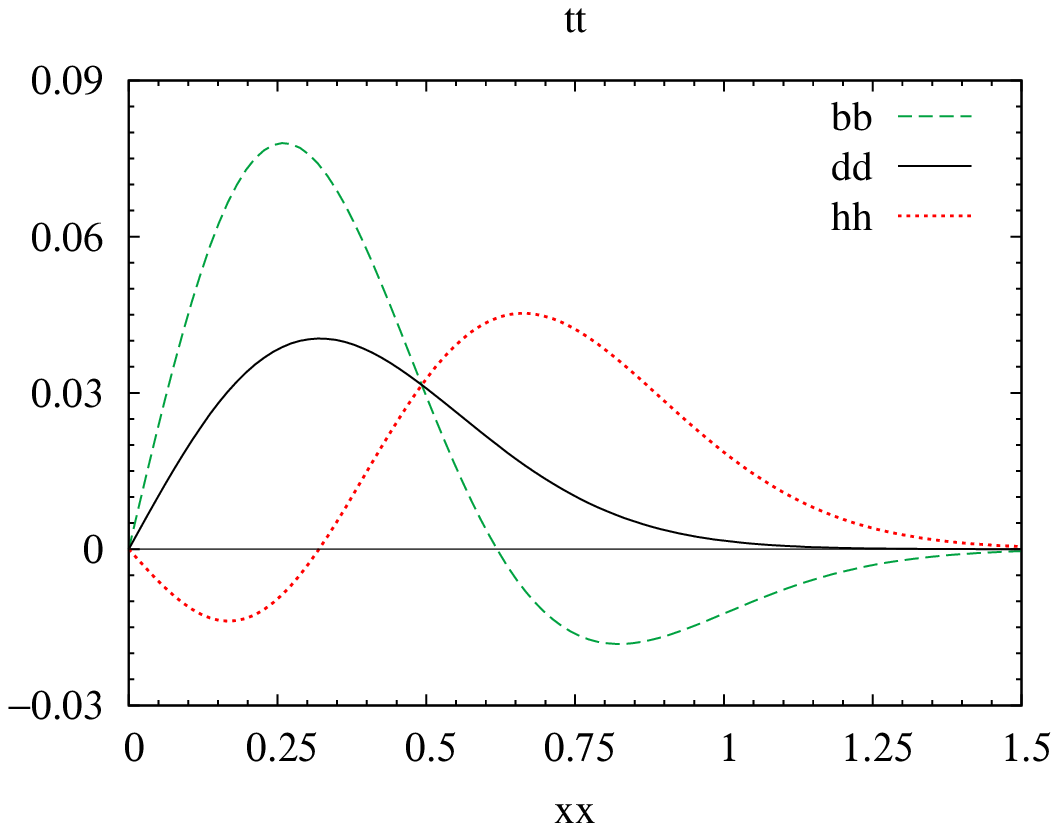}
  \psfrag{xx}[c][c]{\small $x$}
  \psfrag{tt}[c][c]{\small $x\ms (H^{s} - H^{\bar{s}})$}
  \psfrag{bb}[l][c]{\footnotesize $A=0\quad           \bar{A}=4$}
  \psfrag{dd}[l][c]{\footnotesize $\phantom{A}=0\quad \phantom{A}=0$}
  \psfrag{hh}[l][c]{\footnotesize $\phantom{A}=4\quad \phantom{A}=0$}
\hspace{0.2em}
\includegraphics[width=0.49\textwidth,%
  bb=80 50 385 295]{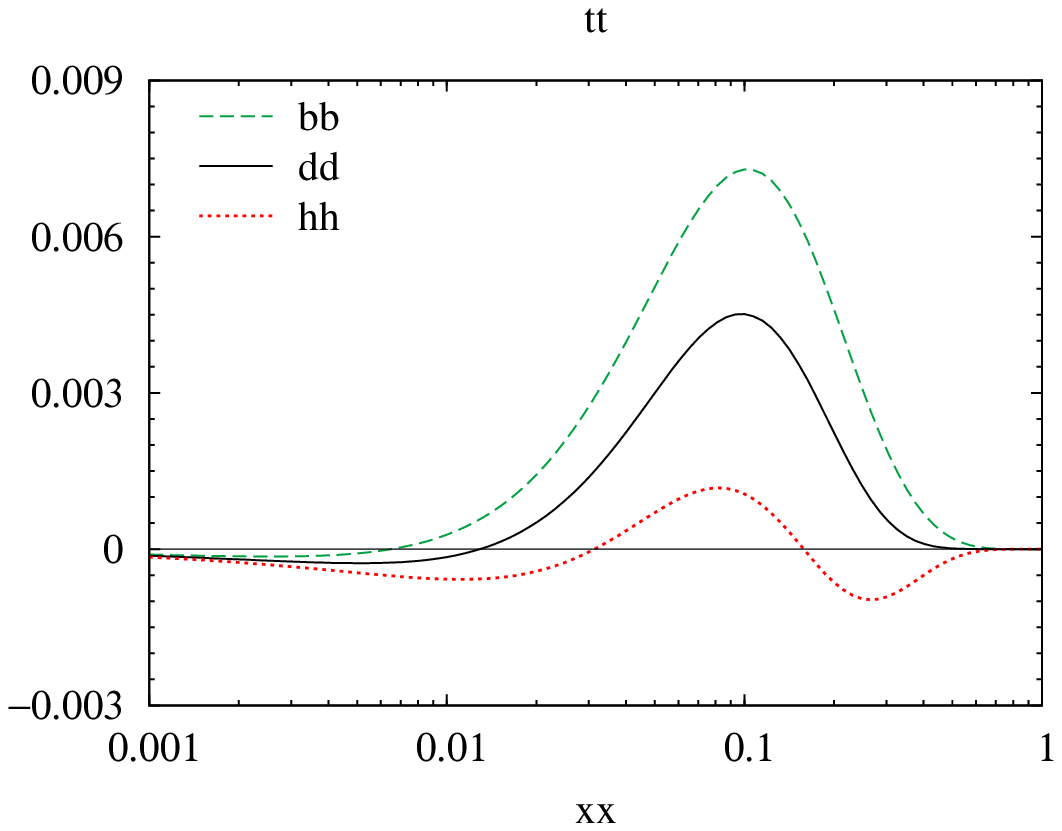}
\end{center}
\caption{\label{fig:alt-bx} Left: the difference in the impact
  parameter distributions for strange quarks and antiquarks at
  $x=0.2$.  The distributions are multiplied with $2\pi b$, so that
  the area under each curve gives $s(x) - \bar{s}(x)$.  Right: the
  integrand \protect\eqref{F1s-int} of the strange Dirac form factor
  at $-t= 0.5\gev^2$.  The curves are for the ansatz in
  \protect\eqref{alt-ansatz} and \eqref{alt-prof} with $\alpha'=
  1\gev^{-2}$ and different values of $A$ and $\bar{A}$ (given in
  units of $\gev^{-2}$).  For the parton densities we take CTEQ6.5S
  set $-0$ at $\mu= 2\gev$.}
\end{figure}

It is however important to realize that at $\mu= 2\gev$, where we
formulate our model, the individual distributions of $s$ and $\bar{s}$
are not valence-like as they would be in a model valid at low
resolution scale.  For sets $-0$, $-1$ and $-2$ of the CTEQ6.5S
parameterization at $\mu= 2\gev$ we find
\begin{equation}
  \label{s+sbar_small-x}
x (s+\bar{s}) \approx 0.2\, x^{-0.2}
\end{equation}
in the region $10^{-5} < x < 10^{-4}$, which is to be compared with
\eqref{s-sbar_small-x}.  For $x < 10^{-2}$ the ratio $(s-\bar{s})
/(s+\bar{s})$ does not exceed $1\%$ in absolute size.  It is natural
to assume that the bulk of $s$ and $\bar{s}$ in that region is
generated through gluon splitting $g\to s\bar{s}$ as described by
perturbative evolution.  This mechanism does not introduce an
asymmetry in the transverse distribution of $s$ and $\bar{s}$.  When
introducing a more general ansatz for $H^s - H^{\bar{s}}$ than
\eqref{master-ansatz} we should hence make sure that the strong
cancellation between $s$ and $\bar{s}$ at small $x$ takes place not
only in the forward limit but also at nonzero $t$.  With this in mind,
we explore a variant of \eqref{master-ansatz}, given by
\begin{equation}
  \label{alt-ansatz}
H^s(x,t) - H^{\bar{s}}(x,t)  = 
  s(x)\, e^{\ms t f_s(x)} - \bar{s}(x)\, e^{\ms t \bar{f}_{s}(x)}
\end{equation}
with
\begin{align}
  \label{alt-prof}
f_s(x) &= \alpha' (1-x) \log\frac{1}{x} + A\ms x (1-x)^2 \,,
\nonumber \\
\bar{f}_{s}(x) &=
  \alpha' (1-x) \log\frac{1}{x} + \bar{A}\ms x (1-x)^2 \,,
\end{align}
where the prefactor $x$ in front of $A$ and $\bar{A}$ guarantees the
cancellation just discussed, as long as $-t A$ and $-t \bar{A}$ are
not too large.  In the following we take values $2\gev^{-2}$ and
$4\gev^{-2}$ for either $A$ or~$\bar{A}$.  With $\alpha'= 1\gev^{-2}$
this respectively
corresponds to a change of $f_s(x)$ or $\bar{f}_s(x)$ by a factor
$1.2$ and $1.4$ at $x=0.2$, which one may view as a typical momentum
fraction for $s$ and $\bar{s}$ in a model at low scale, where
nonperturbative effects could generate an asymmetric distribution in
impact parameter.  In the left panel of Fig.~\ref{fig:alt-bx} we show
the corresponding impact parameter densities $s(x,b) - \bar{s}(x,b)$
and see that they indeed have nodes in $b$ when $A$ or $\bar{A}$ is
equal to $4\gev^{-2}$.

\begin{figure}
\begin{center}
  \psfrag{xx}[c][c]{\small $-t\; [\gev^2]$}
  \psfrag{tt}[c][c]{\small $F_1^s(t)$}
  \psfrag{bb}[r][c]{\footnotesize $A=0\quad           \bar{A}=4$}
  \psfrag{bd}[r][c]{\footnotesize $\phantom{A}=0\quad \phantom{A}=2$}
  \psfrag{dd}[r][c]{\footnotesize $\phantom{A}=0\quad \phantom{A}=0$}
  \psfrag{dh}[r][c]{\footnotesize $\phantom{A}=2\quad \phantom{A}=0$}
  \psfrag{hh}[r][c]{\footnotesize $\phantom{A}=4\quad \phantom{A}=0$}
\includegraphics[width=0.49\textwidth,%
  bb=80 50 385 295]{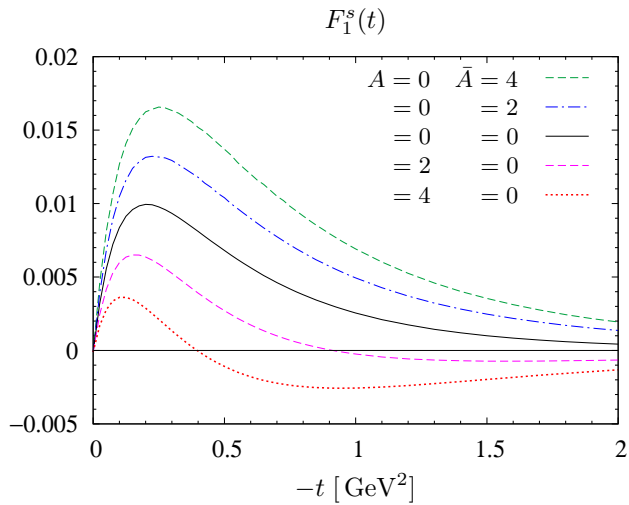}
\end{center}
\caption{\label{fig:alt-F1s} The strange form factor obtained with the
  ansatz specified in Fig.~\protect\ref{fig:alt-bx}.  The curve for
  $A= \bar{A} =0$ corresponds to our default prediction in the
  previous subsections.}
\end{figure}

The form factors obtained with this ansatz are shown in
Fig.~\ref{fig:alt-F1s}.
For $\bar{A} >0$, where $\bar{s}$ is concentrated at larger impact
parameters than $s$, we find that $F_1^s$ is increased in size but
not much changed in shape compared with our default prediction with
$A= \bar{A} =0$.  In contrast, we find that for sufficiently large $A
> 0$ the form factor $F_1^s$ changes sign at some finite $-t$.  We can
understand this at the level of the form factor integrand: for $f_s(x)
> \bar{f}_s(x)$ the exponential factors in \eqref{alt-ansatz} give a
stronger suppression in the first term, so that at large enough $x$
and $-t$ one can have $H^s(x,t) < H^s(x,t)$ despite $s(x) >
\bar{s}(x)$.  This is illustrated in the right panel of
Fig.~\ref{fig:alt-bx}.

As discussed above, the meson cloud picture suggests that $s$ has
smaller rather than larger typical impact parameters than $\bar{s}$,
so that we do not see a particular physics motivation for our examples
with $A >0$.  However, they show that certain nontrivial correlations
between the $x$ and $b$ dependence of the $s$ and $\bar{s}$
distributions can have quite drastic effects on $F_1^s(t)$, which may
be observable in experiments with sufficient sensitivity and kinematic
coverage.


\subsection{Other form factors}

With our model \eqref{master-ansatz} for $H^s - H^{\bar{s}}$ we can
also evaluate the Mellin moment $A_{3,0}^s(t) = \int_{-1}^1 dx\, x^2
H^s(x,t)$, which is a form factor of an operator with two covariant
derivatives between the strange quark field and its conjugate.  The
factor $x^2$ strongly suppresses small $x$ values, and with the CTEQ
parameterizations for $s-\bar{s}$ the integral is dominated by an $x$
region where the integrand has a definite sign.  As a consequence
$|A_{3,0}^s(t)|$ decreases monotonically with $-t$.  Its value at
$t=0$ is tiny, ranging from $-10^{-4}$ to $4 \times 10^{-4}$ for the
CTEQ6 and CTEQ6.5S parameterizations at $\mu= 2\gev$.  Exceptions are
the values $-6 \times 10^{-4}$ and $10^{-3}$ for CTEQ6B$-$ and
CTEQ6B$+$, respectively.

We do not attempt here to model the strangeness contributions to the
energy-momentum and axial form factors, $A_{2,0}^s(t)$ and $F_A^s(t)$,
which according to \eqref{A20} and \eqref{axial} are respectively
related to $s+\bar{s}$ and $\Delta s + \Delta\bar{s}$.  These
distributions mix with gluons under evolution, which invalidates
simple ans\"atze based on Regge trajectories for their small-$x$
behavior.  This also holds at finite $t$ \cite{Diehl:2007}.  Since
even the $t$ dependence of $H^u + H^{\bar{u}}$ and $H^d + H^{\bar{d}}$
is barely constrained at present, we see no clear guidance for how to
model profile functions of $H^s + H^{\bar{s}}$ and $\widetilde{H}^s +
\widetilde{H}^{\bar{s}}$.  We expect however that these distributions
have no zeroes in $x$, which certainly holds for their values at $t=0$
according to current PDF parameterizations.  We therefore predict the
corresponding form factors to decrease monotonically in absolute size,
with values at $t=0$ given by the moments in Tables~\ref{tab:moments}
and \ref{tab:pol-mom}.


\section{Summary}
\label{sec:sum}

We have discussed several measures of strangeness in the nucleon.  
Strange quarks and antiquarks are not particularly rare in the proton:
their contribution $\mom{x (s+\bar{s})}$ to the nucleon momentum is
only suppressed by about a half compared with the one from light flavor
antiquarks, $\mom{x (\bar{u}+\bar{d})}$.  Their contribution
$\mom{\Delta s + \Delta\bar{s}}$ to the spin of the proton is not well
determined at present, but there are no indications that it is very
much suppressed compared with $\mom{\Delta\bar{u}+\Delta\bar{d}}$.

More subtle quantities are \emph{asymmetries} between strange quarks
and antiquarks, most notably the asymmetry between the parton
densities $s(x)$ and $\bar{s}(x)$, and the strange Dirac and Pauli
form factors $F_1^s(t)$ and $F_2^s(t)$.  A two dimensional Fourier
transform of $F_1^s(t)$ yields the difference of spatial distributions
for $s$ and $\bar{s}$ in the transverse plane, whereas
$s(x)-\bar{s}(x)$ gives the difference of their distribution in
longitudinal momentum.  The two asymmetries are connected via
generalized parton distributions at zero skewness, for which we have
made a model ansatz in order to explore this connection
quantitatively.  Using as an input different sets of $s-\bar{s}$
distributions extracted by the CTEQ Collaboration, we find values of
$F_1^s\bigl( t\simeq -0.1\gev^2 \bigr)$ between $-0.006$ and $0.012$,
in good agreement with current experimental constraints.  Many
theoretical analyses of the electromagnetic nucleon form factors
neglect the strangeness contributions.  With our estimates this is at
most a $3\%$ effect for $F_1^p(t)$.  However, $|F_1^s(t)|$ might
amount to as much as $1/6$ of $F_1^n(t)$ at $-t= 0.2\gev^2$.  For
higher $-t$ the relative contribution quickly decreases, whereas for
lower $-t$ it may even be larger.

The general features of our model ansatz for generalized parton
distributions lead to correlations between the $x$ dependence of
$s-\bar{s}$ and the shape of $F_1^s(t)$.  The best fits in the PDF
extractions \cite{Lai:2007dq,Thorne:2007de,Olness:2003wz} yield forms
where $s-\bar{s}$ is negative for small $x$ and positive for large
$x$.  With our ansatz, this gives a negative derivative $d F_1^s(t)
/dt$ at $t=0$ and a maximum of the form factor at some value of $-t$.
With a zero crossing of $s-\bar{s}$ at $x$ between $10^{-2}$ and
$10^{-1}$, we find this maximum at $-t$ between $0.2\gev^2$ and
$0.4\gev^2$.  Analogous correlations are seen to hold between the
combination $\frac{2}{3} (d-\bar{d}) - \frac{1}{3} (u-\bar{u})$ of
valence quark distributions and the neutron form factor $F_1^n(t)$,
which we take as support for our predictions.  Finally, a rapid
decrease of $s-\bar{s}$ with $x$ reflects itself in a faster decrease
of $F_1^s(t)$ compared with $F_1^n(t)$ for large $-t$.  It will be
interesting to confront these predictions with future data from parity
violating electron-nucleon scattering.


\section*{Acknowledgments}

We gratefully thank H.-W. Hammer, R. Sassot, R. Thorne, A. Thomas,
Wu-Ki Tung and R. Young for valuable discussions and correspondence.
This work is supported by the Integrated Infrastructure Initiative of
the European Union under contract number RII3-CT-2004-506078.
T.F.\ is supported by the German Ministry of Research (BMBF) under
contract No.~05HT6PSA.  He is grateful to the Physics Department and
the group of Andrzej Buras at the Technical University in Munich for
the warm hospitality extended to him in the final stage of this work,
as well as for financial support from the Cluster of Excellence
``Origin and Structure of the Universe''.


\end{document}